\documentclass{aa}  
\usepackage{graphicx}
\usepackage{color}
\usepackage{longtable}
\usepackage{subfig,float}
\usepackage{txfonts}
\newcommand{\pnh}{\object{Hen\,2$-$39}\xspace}
\newcommand{\pngh}{\object{PN\,G283.8$-$04.2}\xspace}
\newcommand{\Msol}{$M_\odot$\xspace}
\newcounter{Rco}

\newcommand{\Ionst}[1]{\setcounter{Rco}{#1}\Roman{Rco}}
\newcommand{\Ion}[2]{\mbox{#1\,{\scriptsize\Ionst{#2}}}}

\newcommand{\Ionww}[3]{\mbox{#1\,{\scriptsize\Ionst{#2}}~$\lambda\lambda\,#3$\,\AA}\xspace}

\newcommand{\Jonww}[3]{\mbox{\ion{#1}{#2}~$\lambda\lambda\,#3$\,\AA}\xspace}

\newcommand{\logg}{\mbox{$\log g$}\xspace}
\newcommand{\loggw}[1]{\mbox{$\log g\hspace{-0.5mm} =\hspace{-0.5mm}  #1$}}

\newcommand{\ab}[1]{\mbox{Fig.\,\ref{#1}}}
\newcommand{\sA}[1]{\mbox{(Fig.\,\ref{#1})}}

\newcommand{\sK}[1]{\mbox{(Sect.\,\ref{#1})}}

\newcommand{\ta}[1]{\mbox{Table\,\ref{#1}}}
\newcommand{\sT}[1]{\mbox{(Table\,\ref{#1})}}
\newcommand{\Teff}{\mbox{$T_\mathrm{eff}$}\xspace}
\newcommand{\Teffw}[1]{\mbox{$\Teff\hspace{-0.5mm} =\hspace{-0.5mm} #1 \,\mathrm{K}$}}
\newcommand{\ebv}{\mbox{$E_\mathrm{B-V}$}}
\newcommand{\ebvw}[1]{\mbox{$\ebv\hspace{-0.5mm} =\hspace{-0.5mm} #1$}}

\begin{document}

   \title{Spectral analysis of the barium central star \\
   of the planetary nebula Hen\,2$-$39\thanks{Based on data products from observations made with ESO 
Telescopes at the La Silla Paranal Observatory
under program ID 093.D-0332(A). 
          }}

   \author{L. L\"obling
          \inst{1,2}
          \and
          H.~M.~J. Boffin\inst{1}
          \and 
          D.~Jones\inst{3,4}
          }

   \institute{European Southern Observatory, Karl-Schwarzschild-Str\@. 2, D-85748 Garching bei M\"unchen, Germany
           \and
           Institute for Astronomy and Astrophysics,
           Kepler Center for Astro and Particle Physics,
           Eberhard Karls University,\\
           Sand 1,
           D-72076 T\"ubingen,
           Germany\\
           \email{loebling@astro.uni-tuebingen.de}
           \and 
           Instituto de Astrof\'{i}sica de Canarias, 
           E-38205 La Laguna, Tenerife, Spain
           \and 
           Departamento de Astrof\'{i}sica, Universidad de La Laguna, 
           E-38206 La Laguna, Tenerife, Spain
             }

   \date{Received 19 October 2018; accepted 6 February 2019}

  \abstract
   {Barium stars are peculiar red giants characterized by an overabundance of the elements synthesized in the slow neutron-capture nucleosynthesis (s-process elements) along with an enrichment in carbon. These stars are discovered in binaries with white dwarf companions. The more recently formed of these stars are still surrounded by a planetary nebula.}
   {Precise abundance determinations of the various s-process elements, of further key elements that act as indicators for effectiveness of nucleosynthesis on the asymptotic giant branch and, especially, of the lightest, short-lived radionuclide technetium will establish constraints for the formation of s-process elements in asymptotic giant branch stars as well as mass transfer through, for example,  stellar wind, Roche-lobe overflow, and common-envelope evolution.}
   {We performed a detailed spectral analysis of the K-type subgiant central star of the planetary nebula \pnh based on high-resolution optical spectra obtained with the Ultraviolet and Visual Echelle Spectrograph at the Very Large Telescope using local thermodynamic equilibrium model atmospheres.}
   {We confirm the effective temperature of $\Teff = (4350 \pm 150)$\,K for the central star of the planetary nebula \pnh. It has a photospheric carbon enrichment of $[\mathrm{C}/\mathrm{H}]= 0.36 \pm \textcolor{black}{0.08}$ and a barium overabundance of $[\mathrm{Ba}/\mathrm{Fe}]= \textcolor{black}{1.8 \pm 0.5}$. We find a deficiency for most of the iron-group elements (calcium to iron) and establish an upper abundance limit for technetium ($\log \epsilon_\mathrm{Tc} < 2.5$).
 }
   {The quality of the available optical spectra is not sufficient to measure abundances of all s-process elements accurately. {Despite large uncertainties on the abundances as well as on the model yields, the derived abundances are most consistent with a progenitor mass in the range 1.75-3.00\,\Msol} and a metallicity of
$[\mathrm{Fe}/\mathrm{H}]= -\textcolor{black}{0.3 \pm 1.0}$. 
This result leads to the conclusion that the formation of such systems requires a relatively large mass transfer that is most easily obtained via wind-Roche lobe overflow. }

\titlerunning{Spectral analysis of the Ba CSPN of \pnh}
  \maketitle
    
%

\section{Introduction \label{sect:intro}}

   So far, only a small number of planetary nebulae (PNe) have been identified to host a binary with a giant or subgiant component dominating the optical wavelength range and showing peculiar surface element abundances that indicate late stage stellar evolution nuclear synthesis. These stars exhibit signatures of slow neutron-capture nucleosynthesis (s-process) in their spectra and in some cases an enrichment in carbon (C). 
   
For the object of this work, the central star (CS) of the PN \pnh 
\citep[\pngh, \object{Wray\,16$-$64}; ][]{henize1967,ackeretal1992,wray1966}, \cite{miszalskietal2013hen} determined an overabundance for the s-process element barium (Ba) of $[\mathrm{Ba}/\mathrm{Fe}]\footnote{$[\mathrm{A}/\mathrm{B}] = \log (n_\mathrm{A} / n_\mathrm{B} ) - \log (n_{\mathrm{A},\odot} / n_{\mathrm{B},\odot} )$ with the number fractions $n$ for element A and B}=1.5 \pm 0.25$ and an enrichment of $[\mathrm{C}/\mathrm{H}]= 0.42 \pm 0.02$ in a spectral analysis based on mid-resolution spectra obtained with the Southern African Telescope \citep[SALT;][]{buckleyetal2006} with the Robert Stobie Spectrograph \citep[RSS;][]{burghetal2003,kobulnickyetal2003}.
These findings confirm the membership of the K-type nucleus of \pnh in the small group of Ba \textcolor{black}{central stars of planetary nebulae} (CSPNe) along with \object{LoTr\,5}, \object{WeBo\,1}, and \object{Abell\,70} \citep{theveninetal1997,bondetal2003,miszalskietal2012,tyndall13,aller18}. 

   Ba CSPNe are prime examples of progenitors of Ba stars that were described by \cite{bidelmanetal1951}. {Because of their evolutionary status, namely still being on the main sequence or a red giant}, these stars did not yet experience AGB nucleosynthesis and, thus, cannot have synthesized \textcolor{black}{heavy} elements. \textcolor{black}{\cite{mcclureetal1980} discovered the binary nature of Ba stars and proposed that mass transfer was key to explain these sources}. \cite{boffinetal1988} \textcolor{black}{performed detailed simulations of wind mass transfer to explain} the pollution of the Ba star from an evolved companion with the products of asymptotic giant branch (AGB) nucleosynthesis that are dredged up to the stellar surface \citep{herwig2005,wernerherwig2006}. More recently, other mechanisms were proposed in which the material is transferred to the still unevolved companion \citep{boffinetal2015} via Roche-lobe overflow \citep[RLOF; e.g., ][]{hanetal1995} or wind-RLOF \citep[e.g., ][]{nagaeetal2004,mohamedetal2007,abateetal2013}.
   
   This scenario is strongly supported by the fact that, so far, all Ba stars are found in binaries with white dwarf (WD) companions \citep{mcclureetal1980,mcclure1983,jorissenetal1988,mcclure1990,jorissenetal1998}, which is also definitely clear for the Ba CSPNe  since the Ba star is not hot enough to ionize the ambient ejected material that is visible as the surrounding PN. Although the (pre-)WD companion must be there without any doubt, it can be difficult to detect against the bright companion even in the UV. Recently, more and more WD companions of Ba stars \citep{grayetal2011} and pre-WD companions of Ba CSPNe have been detected \citep[e.g., Abell\,70, ][]{miszalskietal2012}, which doubtlessly confirms the formation scenario. The still poorly understood mechanism of mass transfer in these systems is subject of ongoing research. The challenge is to determine their orbital parameters, such as eccentricity and period and to reproduce these with theoretical binary evolution models \citep{saladinoetal2018,demarco2009}.
   
   These stars are expected to show orbital periods of several hundred days, which are typical values for Ba stars \citep{jorissenetal1998}. However, there is the CSPN binary in the Necklace Nebula \citep[PN~G054.6$-$03.4; ][]{corradietal2011} standing out toward shorter periods. \cite{miszalskietal2013} found a period of 1.16\,d for the post-CE system from the analysis of the C-dwarf secondary. On the other side of the period range, current analyses also indicate that there are systems with values up to several years and with larger eccentricities \citep{jonesetal22017}. 

   Ba CSPNe are ideal to study AGB nucleosynthesis. They provide a snapshot of an evolutionary stage with ideal conditions for analyzing not only the polluted cool (sub)giant star but also the ejected material of the nebula around the polluting post-AGB star \citep[e.g., ][]{madonnaetal2017,madonnaetal2018}. 
The short duration of the PN phase ($\approx 10^4$\,yrs) guarantees that the mass transfer happened recently and that the companion has not yet had time to adjust. Also, in some cases, the polluted star is still unevolved and did not experience the first dredge-up (DU) that would affect the surface element composition including the nucleosynthesis outcomes from the polluting post-AGB star.

   By comparing the results of \textcolor{black}{our} comprehensive spectral analysis to theoretical AGB nucleosynthesis models \citep{karakasetal2016,karakasetal2018}, new insights into Ba stars and PNe are gained. It is worth mentioning the s-process mechanisms including atomic reaction rates, the source of neutrons and neutron exposure, internal stellar structures, and mixing processes occurring in a thermal-pulsing AGB star. This allows us to constrain the progenitor mass of the post-AGB star and the number of thermal pulses (TPs) on the AGB. Including binary evolution models \citep{saladinoetal2018,demarco2009}, these objects offer the opportunity to study the CE process and (wind-)RLOF, which are still far from being understood \citep{miszalskietal2013,jonesetal2017}, and in addition the fraction of mass transferred \citep{boffinetal1988} and, following from this, the dilution factor in the Ba star itself and, thus, the mixing processes at work in (sub)giant stars \citep{hustietal2009}. In particular, Ba CSPNe such as \pnh offer the possibility to detect technetium (Tc), which is the lightest element with no stable isotopes, in their atmospheres. This element was first detected by \cite{merrill1952} in the atmospheres of red giants, which proved that it is synthesized in evolved stars, since the half-life of $^{99}$Tc of 210\,000\,yrs\footnote{Los Alamos National Laboratory Periodic Table \url{http://periodic.lanl.gov}} is much shorter than the previous giant evolutionary phase. It is thus only observed in AGB stars currently undergoing thermal pulses \citep[TPs; ][]{vanecketal1999,lebzelteretal2003} and, hence, the determination of the Tc surface abundance of the Ba CSPN indicates the mass-transfer link between the binary components in the PN and establishes a definite indicator for the existence of the third dredge-up (TDU). 
   Assuming a typical post-AGB age of some $10^3-10^4$\,yrs \citep{millerbertolami2016} for the primary component and taking into account that the dynamical process of mass transfer is short compared to this number \citep{ibenetal1993,chenetal2017}, a large fraction of the transferred Tc should still be present in the stellar atmosphere.
   
   We describe the observations, stellar atmosphere models, and analysis techniques in Sect.~\ref{sect:obs} and \ref{sect:atmos}, respectively.
   The spectral analysis follows in Sects.~\ref{sect:stellarpara} and \ref{sect:abund}. The results are discussed in Sect.~\ref{sect:discussion}. We summarize and conclude in Sect.~\ref{sect:rescon}.


\section{Observations\label{sect:obs}}

n2-39The spectral analysis of \pnh is based on spectra in the optical wavelength range obtained with the Ultraviolet and Visual Echelle Spectrograph \citep[UVES;][]{dekker00} at the Very Large Telescope (VLT) at the Paranal Observatory of the European Southern Observatory under ESO program 093.D$-$0332(A). The data products created from this data were retrieved from the ESO Science Archive Facility. The observation log, \textcolor{black}{including the signal-to-noise ratio (S/N)}, of the spectra used in this paper is shown in Tab.~\ref{tab:obslog}. All spectra were taken with a resolving power of $R = 42\,000 - 44\,000$.

We used the spectral analysis code ISpec \citep{blancoetal2014} to determine the radial velocities for each single observation via cross-correlation with a model template spectrum created using the \textcolor{black}{fundamental parameters \Teff, \logg, C abundance, and metallicity} determined by \cite{miszalskietal2013hen}. The heliocentric corrected radial velocities for the 18 observations are given in \ta{tab:vrad}.
To improve the S/N, all observations were shifted to the rest-frame velocity and subsequently co-added. To simulate the resolution of the instrument, all synthetic spectra shown in this work were convolved with Gaussians \textcolor{black}{(full width half maximum $\mathrm{(FWHM)} = 0.12\,{\AA}$)}.

   \begin{table}
   \centering
      \caption[]{Observation time and heliocentric radial velocities for the 18 observations of the Ba CSPN of \pnh.}
         \label{tab:vrad}
         \begin{tabular}{l c}
            \hline
            \hline
            \noalign{\smallskip}
            MJD  & $v_\mathrm{rad}$  \\
            \noalign{\smallskip}
            \hline
            \noalign{\smallskip}
56750.117 &  $44.79 \pm 1.06$ \\
56750.135 &  $44.76 \pm 1.01$ \\
56750.154 &  $46.61 \pm 1.00$ \\
56750.172 &  $45.64 \pm 1.09$ \\
56751.032 &  $46.05 \pm 0.93$ \\
56751.050 &  $46.47 \pm 0.92$ \\
56751.070 &  $46.34 \pm 1.07$ \\
56751.088 &  $46.29 \pm 1.03$ \\
56751.109 &  $45.35 \pm 1.05$ \\
56751.127 &  $45.17 \pm 1.05$ \\
56760.108 &  $45.43 \pm 0.94$ \\
56760.126 &  $45.56 \pm 1.06$ \\
56762.018 &  $45.79 \pm 1.09$ \\
56762.035 &  $45.21 \pm 1.00$ \\
56762.053 &  $45.64 \pm 1.06$ \\
56762.070 &  $45.91 \pm 1.08$ \\
56762.088 &  $45.96 \pm 1.10$ \\
56762.106 &  $46.24 \pm 1.09$ \\
\noalign{\smallskip}
            \hline
         \end{tabular}
   \end{table}

\section{Model atmospheres, atomic data, and analysis techniques\label{sect:atmos}}

We used the stellar synthesis code SPECTRUM\footnote{\url{http://www.appstate.edu/~grayro/spectrum/spectrum.html}} \citep[][version 2.76]{grayetal1994} to calculate synthetic spectra for the analysis of the observed high-resolution spectra with the ATLAS9 model atmosphere grids\footnote{\url{http://kurucz.harvard.edu/grids.html}} \citep{kurucz1991,castellietal2003} as input. These one-dimensional models are based upon the solar abundances from \cite{grevesseetal1998} and are calculated under the presumption of plane-parallel geometry and local thermodynamic equilibrium (LTE), which is valid for {stars in this temperature and gravity regime }\citep{hubenyetal2003}.
For the wavelength values and oscillator strengths of the lines selected in our analysis, we used the values provided within the distribution of SPECTRUM. Data for Tc\,\textsc{i} were retrieved from the Atomic Spectra Database\footnote{\url{https://www.nist.gov/pml/atomic-spectra-database}} of the National Institute of Standards and Technology (NIST). For Tc\,\textsc{ii}, we used the data provided by \cite{palmerietal2007}.
We calculated an extensive \textcolor{black}{grid of synthetic spectra} spanning from $\Teff = 3500$\,K to $6000$\,K ($\Delta \Teffw{250}$ between 4000\,K and 5000\,K and $\Delta \Teffw{500}$ otherwise) and from $\log (g\,/\,\mathrm{cm/s^2}) = 0.0$ to $4.0$ ($\Delta \loggw{0.5}$) with a metallicity of $\mathrm{[M/H]} = -0.3$ around the literature values of $\Teff = 4250 \pm 150$\,K and $\logg = 2.0 \pm 0.5$ \citep{miszalskietal2013hen}. For the determination of abundances, we relied on the model with $\Teff = 4250$\,K and $\logg = 2.5$ and varied the abundance of one single element over a range of at least 2.5\,dex with a step of 0.5\,dex. {The exceptions to this are C, for which we varied the abundance in steps of 0.05\,dex over a range of 0.25\,dex, and N with a range of 1.5\,dex and steps of 0.3\,dex.} 
Since the spectrum is crowded with absorption lines that are broadened by rotation, we could not measure equivalent widths to determine the fundamental parameters. We performed the analysis of the different parameters by selecting wavelength regions that show a strong influence of these particular species. The final values were then derived using a $\chi^2$-method applied to the \textcolor{black}{synthetic spectra} grid for the selected regions.

\section{Stellar parameters \label{sect:stellarpara}}

\subsection{Rotation \label{subsec:rotation}}

To determine $v_\mathrm{rot} \sin i = 38 \pm 5\,\mathrm{km}/\mathrm{s}$, we used a fit of a synthetic spectrum calculated with the literature values given by \cite{miszalskietal2013hen} and convolved with rotational profiles for values from $v_\mathrm{rot} \sin i = 0$ to $50\,\mathrm{km}/\mathrm{s}$ to two regions spanning from $6440$ to $6515$\,{\AA} and from $7030$ to $7070$\,{\AA} dominated by strong C$_2$ molecular absorption bands \sA{fig:rotation}.

\begin{figure} 
  \resizebox{\hsize}{!}{\includegraphics{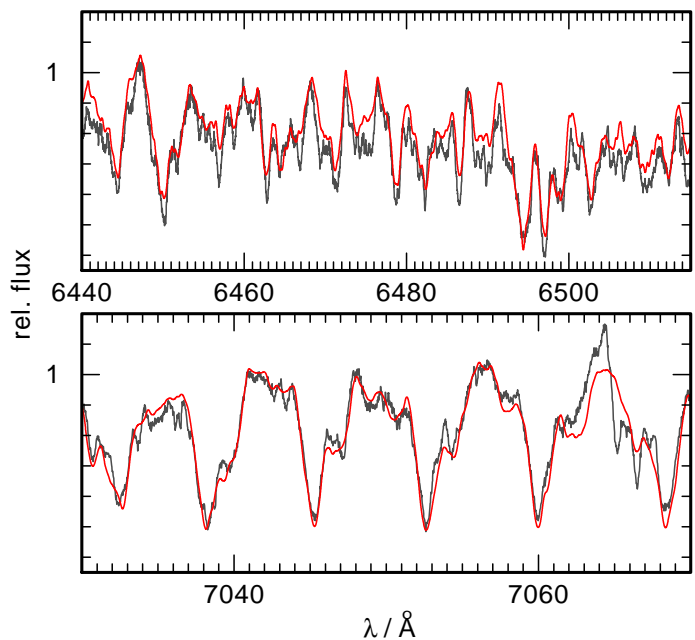}}
   \caption{Synthetic spectra (red) convolved with a rotational profile of $v_\mathrm{rot} = 38\,\mathrm{km}/\mathrm{s}$ 
            around 
            strong C$_2$ and CN molecular absorption bands
            compared with observations (gray).
            }
   \label{fig:rotation}
\end{figure}

\subsection{Effective temperature and surface gravity \label{subsec:tefflogg}}

The many rotationally broadened lines in the observation also hamper the determination of equivalent widths. Thus, we used a set of diagnostic lines of \Ion{Fe}{1}, \Ion{Fe}{2}, \Ion{Ti}{1}, \Ion{Ti}{2}, \Ion{Sc}{1}, \Ion{Sc}{2}, and \Ion{Mg}{1} and performed a $\chi^2$-fit for selected wavelength regions. The set is composed of the lines that \cite{taberneroetal2018} used for a similar spectral analysis. Furthermore, we included some \Ion{Fe}{1} and \Ion{Fe}{2} lines that are used by the Brussels Automatic Code for Characterizing High accuracy Spectra \citep[BACCHUS\footnote{\url{http://www.astro.ulb.ac.be/pmwiki/Spectro/Bacchus}}; ][]{masseronetal2016}. Finally, we added lines for \Ion{Fe}{1}, \Ion{Fe}{2}, \Ion{Ti}{1}, \Ion{Ti}{2}, \Ion{Sc}{1}, and \Ion{Sc}{2} for which we obtained the largest theoretical equivalent widths in the calculation of the synthetic spectra and that did not saturate. The resulting collection of wavelength regions used for the analysis and the list of diagnostic lines is shown in \ta{tab:charlines}. They cover a wide range of different excitation potentials and oscillator strengths. After a first determination of \Teff and \logg, we found a degeneracy in \logg due to the fact that the strength of the computed lines in the regime around the literature values varies very little as a function of \logg for a fixed value of \Teff. 
A spectroscopic determination of \logg is hampered by uncertain values for the distance and brightness \sK{ssect:spectdist}. Thus, we adopt a value of 
$\logg = 2.5 \pm 0.5$ that is typical for Ba giants of that type \citep[e.g., ][]{decastroetal2016}. This approach seems to be reasonable because a change of $\Delta \logg = 0.5$ only marginally affects the derived abundances compared to the significant statistical errors.
We derive $\Teff = (4350 \pm 150)$\,K.
\textcolor{black}{\ab{fig:teff} and \ab{fig:logg} illustrate the spectroscopic determination of these parameters by showing the difference due to a variation of \Teff and \logg.}

\section{Element abundances \label{sect:abund}}

We used model atmospheres with $\Teff = 4250$\,K and $\logg = 2.5$ from the grid and performed line-profile fits for the following elements to determine their abundances. The results are given in \ta{tab:abund}. 
\textcolor{black}{In our analysis, we assumed the atomic data to be correct and did not propagate uncertainties on atomic data. The continuum placement uncertainty is also assumed to be small as the continuum placement is shifted during the fitting procedure.}  
To estimate the impact of a varied \Teff and \logg on the determined abundances, we redid part of our analysis with models with \textcolor{black}{varied $\Teff$ between $4000$ and $4500$\,K at $\logg$ varied between $2.0$ and $3.0$}. We also varied the microturbulence velocity by $\pm 2.0$\,km/s, which was kept fixed at $2.0$\,km/s in the initial analysis.
By far, the impact of a change in temperature is the largest. Compared to this error, the variation in \logg and microturbulence velocity become negligible. Furthermore, we investigated the influence of the metallicity of the model atmosphere grid that was chosen for the analysis on the determined abundances and repeated part of the analysis with different input model metallicities between $-0.5$ and 0. 
Raising (lowering) the metallicity by a certain amount results in a Fe abundance that is lower (higher) by roughly 1.3-fold that amount. Consistency is reached for the grid with $\mathrm{[M/H]} = -0.3$ that gives a Fe abundance of $[\mathrm{Fe}/\mathrm{H}] = \textcolor{black}{0.3 \pm 1.0}$.
{In our analysis, we find a C enrichment but cannot determine the O abundance and, thus, use the solar value. The resulting C/O ratio is larger than one. To test whether it is justified to use an O-rich model atmosphere grid (model 1) with solar abundances, we employed the ATLAS9 code and calculated a test model with a C abundance increased to the value found in our analysis (model 2). Furthermore, we calculated a second test model with the increased C abundance and an O abundance reduced by one dex (model 3). We redid the abundance analysis for C and Ba and found the same abundance for C from model 1 and 2. For the model with increased C and decreased O abundance, we get a C abundance that is 0.03\,dex higher. Compared to model 1, the Ba abundance for model 2 is higher by 0.04\,dex and by 0.2\,dex for model 3. The difference between the O-rich model 1 and the C-rich model 2 is shown in \ab{fig:res}. The effect of the model atmosphere on the abundances is in a range that justifies relying on the available model atmosphere grid. Thus, we did not compute a C-rich grid. However, this adds another uncertainty to the abundances.}
\textcolor{black}{We obtain large errors arising from the crowdedness of the observed spectrum that we estimate by detailed line profile fits and evaluation based on the $\chi$-by-eye method. In many cases, this is the main contributor to the total error. It ranges between 0.05\,dex for C and about 0.4\,dex for the light metals up to about 1\,dex for the iron group and trans-iron elements. To take the uncertainties in \Teff and \logg into account, we did this procedure for the corners of the grid stated above ($\Teff=4000$\,K, $\logg = 3.0$ and $4500$\,K , $2.0$). The abundance errors arising from this effect range between 0.03\,dex for C to about 0.5\,dex for the other metals. The total errors given in \ta{tab:abund} are the maximum differences for the abundances that are possible within the error limits of this grid.}

   \begin{table}
   \centering
      \caption[]{Element abundances determined for \pnh in $\log \epsilon = 12 + \log (n_\mathrm{X} / n_\mathrm{H} )$, $[\mathrm{X}/\mathrm{H}] = \log (n_\mathrm{X} / n_\mathrm{H} ) - \log (n_{\mathrm{X},\odot} / n_{\mathrm{H},\odot} )$, and $[\mathrm{X}/\mathrm{Fe}] = \log (n_\mathrm{X} / n_\mathrm{Fe} ) - \log (n_{\mathrm{X},\odot} / n_{\mathrm{Fe},\odot} )$ with the number fraction $n_\mathrm{X}$ for element X.}
         \label{tab:abund}
         \begin{tabular}{l r@{.}l r@{.}l r@{.}l r@{.}l}
            \hline
            \hline
            \noalign{\smallskip}
            Element & \multicolumn{2}{c}{$\log \epsilon$}   &  \multicolumn{2}{c}{$[\mathrm{X}/\mathrm{H}]$} &  \multicolumn{2}{c}{$[\mathrm{X}/\mathrm{Fe}]$} &  \multicolumn{2}{c}{Error}\\
            \noalign{\smallskip}
            \hline
            \noalign{\smallskip}
           
C  & $8$ & $9$ & $ 0$ & $36$ & $ 0$ & $71$ &  \textcolor{black}{$0$} & \textcolor{black}{$08$} \\ 
N  & $8$ & $3$ &  $ 0$ & \color{black}$3$ &  $ 0$ & \color{black}$7$ &  \textcolor{black}{$0$} & \textcolor{black}{$8$} \\ 
Na & $6$ & $0$ &  $-0$ & \color{black}$3$ &  $ 0$ & \color{black}$1$ &  \textcolor{black}{$0$} & \textcolor{black}{$7$} \\ 
Al & $5$ & $8$ &  $-0$ & \color{black}$7$ &  $-0$ & \color{black}$3$ &  \textcolor{black}{$1$} & \textcolor{black}{$0$} \\ 
S  & $8$ & $1$ & $ 0$ &  \color{black}$8$ & $ 1$ &  \color{black}$2$ &  \textcolor{black}{$1$} & \textcolor{black}{$2$} \\ 
K  & $4$ & $8$ & $-0$ &  \color{black}$3$ & $ 0$ &  \color{black}$1$ &  \textcolor{black}{$1$} & \textcolor{black}{$0$} \\ 
Ca & $5$ & $9$ & $-0$ &  \color{black}$4$ & $-0$ &  \color{black}$1$ &  \textcolor{black}{$1$} & \textcolor{black}{$0$} \\ 
Sc & $<2$ & $4$ & $<-0$ &  \color{black}$7$ & $<-0$ &  \color{black}$4$ & \multicolumn{2}{c}{}  \\ 
Ti & $4$ & $0$ & $-0$ &  \color{black}$9$ & $-0$ &  \color{black}$6$ &  \textcolor{black}{$1$} & \textcolor{black}{$4$} \\ 
V  & $3$ & $1$ & $-0$ &  \color{black}$9$ & $-0$ &  \color{black}$6$ &  \textcolor{black}{$1$} & \textcolor{black}{$0$} \\ 
Cr & $5$ & $1$ & $-0$ &  \color{black}$5$ & $-0$ &  \color{black}$2$ &  \textcolor{black}{$1$} & \textcolor{black}{$3$} \\ 
Mn & $5$ & $1$ & $-0$ &  \color{black}$3$ & $ 0$ &  \color{black}$0$ &  \textcolor{black}{$1$} & \textcolor{black}{$0$} \\ 
Fe & $7$ & $1$ & $-0$ &  \color{black}$3$ & \multicolumn{2}{c}{}  &  \textcolor{black}{$1$} & \textcolor{black}{$0$} \\ 
Co & $5$ & $1$ & $ 0$ &  \color{black}$2$ & $ 0$ &  \color{black}$5$ &  \textcolor{black}{$1$} & \textcolor{black}{$0$} \\ 
Ni & $6$ & $5$ & $ 0$ &  \color{black}$2$ & $ 0$ &  \color{black}$6$ &  \textcolor{black}{$1$} & \textcolor{black}{$3$} \\ 
Cu & $5$ & $0$ & $ 0$ &  \color{black}$8$ & $ 1$ &  \color{black}$2$ &  \textcolor{black}{$1$} & \textcolor{black}{$5$} \\ 
Zn & $<5$ & $8$ & $<1$ &  \color{black}$2$ & $<1$ &  \color{black}$6$ & \multicolumn{2}{c}{}  \\ 
Rb & $3$ & $7$ & $ 1$ &  \color{black}$1$ & $ 1$ &  \color{black}$4$ & \textcolor{black}{$1$} & \textcolor{black}{$3$} \\ 
Sr & $3$ & $6$ & $ 0$ &  \color{black}$6$ & $ 1$ &  \color{black}$0$ & \textcolor{black}{$1$} & \textcolor{black}{$5$} \\ 
Y  & $2$ & $3$ & $ 0$ &  \color{black}$0$ & $ 0$ &  \color{black}$4$ & \textcolor{black}{$1$} & \textcolor{black}{$5$} \\ 
Zr & $2$ & $4$ & $-0$ &  \color{black}$2$ & $ 0$ &  \color{black}$2$ & \textcolor{black}{$1$} & \textcolor{black}{$5$} \\ 
Nb & $<2$ & $0$ & $ <0$ &  \color{black}$7$ & $ <1$ &  \color{black}$0$ &  \multicolumn{2}{c}{} \\ 
Mo & $2$ & $9$ & $ 1$ &  \color{black}$0$ & $ 1$ &  \color{black}$4$ & \textcolor{black}{$ 1$} & \textcolor{black}{$3$} \\ 
Tc & $<2$ & $5$ & \multicolumn{2}{c}{}&\multicolumn{2}{c}{} & \multicolumn{2}{c}{}  \\ 
Ru & $<3$ & $5$ & $ <1$ &  \color{black}$7$ & $ <2$ &  \color{black}$1$ &  \multicolumn{2}{c}{} \\ 
Ba & $3$ & $6$ & $ 1$ &  \color{black}$4$ & $ 1$ &  \color{black}$8$ & \textcolor{black}{$0$} & \textcolor{black}{$5$} \\ 
La & $2$ & $3$ & $ 1$ &  \color{black}$1$ & $ 1$ &  \color{black}$5$ & \textcolor{black}{$1$} & \textcolor{black}{$6$} \\ 
Ce & $<3$ & $5$ & $ <2$ &  \color{black}$0$ & $ <2$ &  \color{black}$3$ &  \multicolumn{2}{c}{} \\  
Pr & $<3$ & $0$ & $ <2$ &  \color{black}$4$ & $ <2$ &  \color{black}$7$ &  \multicolumn{2}{c}{} \\ 
Nd & $1$ & $9$ & $ 0$ &  \color{black}$4$ & $ 0$ &  \color{black}$8$ & \textcolor{black}{$1$} & \textcolor{black}{$5$} \\ 
Sm & $<1$ & $7$ & $ <0$ &  \color{black}$8$ & $ <1$ &  \color{black}$1$ &  \multicolumn{2}{c}{} \\ 
Eu & $<1$ & $1$ & $ <0$ &  \color{black}$6$ & $ <1$ &  \color{black}$0$ &  \multicolumn{2}{c}{} \\ 
Gd & $<2$ & $6$ & $ <1$ &  \color{black}$5$ & $ <1$ &  \color{black}$8$ &  \multicolumn{2}{c}{} \\ 
Tb & $<0$ & $8$ & $ <0$ &  \color{black}$5$ & $ <0$ &  \color{black}$8$ &  \multicolumn{2}{c}{} \\ 
Dy & $<4$ & $5$ & $ <3$ &  \color{black}$4$ & $ <3$ &  \color{black}$8$ &  \multicolumn{2}{c}{} \\ 
Er & $<2$ & $4$ & $ <1$ &  \color{black}$6$ & $ <1$ &  \color{black}$9$ &  \multicolumn{2}{c}{} \\ 
Hf & $<1$ & $8$ & $ <1$ &  \color{black}$0$ & $ <1$ &  \color{black}$4$ &  \multicolumn{2}{c}{} \\ 
W  & $1$ & $4$ & $ 0$ &  \color{black}$7$ & $ 1$ &  \color{black}$1$ & \textcolor{black}{$1$} & \textcolor{black}{$5$} \\ 
Os & $<2$ & $9$ & $ <1$ &  \color{black}$5$ & $ <1$ &  \color{black}$8$ &  \multicolumn{2}{c}{} \\ 
\noalign{\smallskip}
            \hline
            
         \end{tabular}
   \end{table}

\paragraph{Carbon.}
We analyzed the C abundance using spectrum synthesis calculations for the region of strong C$_2$ absorption from 4650\,{\AA} to 4737\,{\AA} \sA{fig:c_abund}. We confirm the C enhancement and our result of $[\mathrm{C}/\mathrm{H}] = 0.36 \pm \textcolor{black}{0.08}$ agrees within $1\sigma$ with the value derived by \citet{miszalskietal2013hen} from their mid-resolution spectra.

\begin{figure}[t] 
  \resizebox{\hsize}{!}{\includegraphics{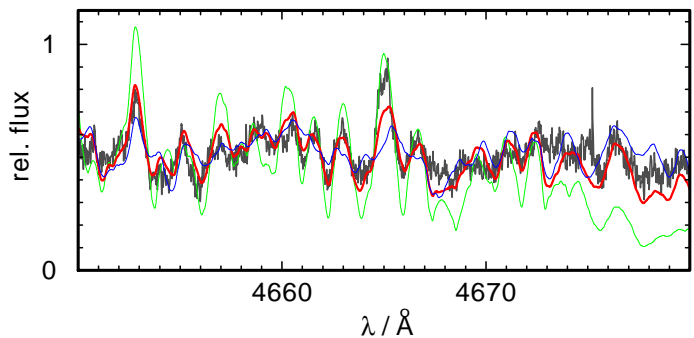}}
   \caption{Observation (gray) of \pnh compared to model spectra \textcolor{black}{with T$_{\rm eff}$=4250~K and $\log g=$2.5} for a selected region of strong C$_2$ absorption for $[\mathrm{C}/\mathrm{H}] = 0.46, 0.36, 0.26$ (green, red, and blue, respectively).
            }
   \label{fig:c_abund}
\end{figure}

\paragraph{Nitrogen.}
Using the C abundances, we derived the N abundance from synthetic calculations for the wavelength regions 7030$-$7070\,{\AA} and 7900$-$8100\,{\AA} affected by strong CN absorption bands. \ab{fig:n_abund} shows the best result. We find N to be enriched to the same level as C with $[\mathrm{N}/\mathrm{H}] = \textcolor{black}{0.3 \pm 0.8} $. 
We could not identify any line of oxygen in the observed spectrum and, thus, were unable to fix an abundance value for O. {In our analysis, we adopt the solar value}.
To get an idea of the $^{12}$C/$^{13}$C ratio, we analyzed the CN absorption band in the region 
8100$-$8200\,{\AA} and included the line list for $^{13}$C$^{14}$N from \citet{snedenetal2014} \sA{fig:c13}. \textcolor{black}{From the inspection of the observation, we cannot claim to find a enhancement in $^{13}$C resulting in $^{12}$C/$^{13}$C lower than the solar value of 90, although this cannot be ruled out.} 

\begin{figure} 
  \resizebox{\hsize}{!}{\includegraphics{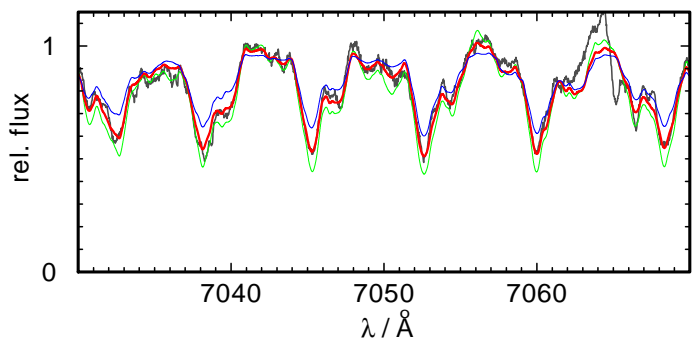}}
   \caption{Like \ab{fig:c_abund}, for strong CN absorption for $[\mathrm{N}/\mathrm{H}] = \textcolor{black}{1.1, 0.3, -0.5}$ (green, red, and blue, respectively).
            }
   \label{fig:n_abund}
\end{figure}

\begin{figure} 
  \resizebox{\hsize}{!}{\includegraphics{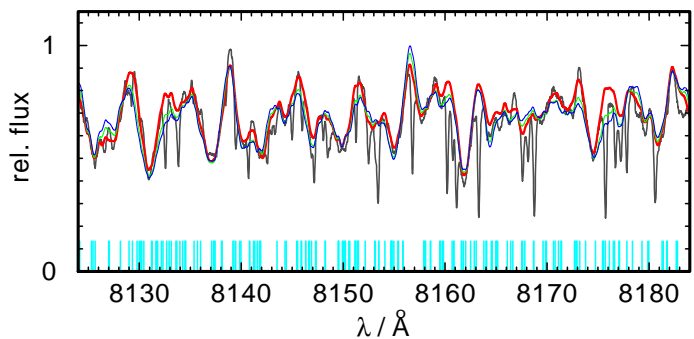}}
   \caption{Like \ab{fig:c_abund}, for strong CN absorption for $^{12}\mathrm{C} / ^{13}\mathrm{C} = 90, 10, 5$ (red, green, and blue, respectively). $^{13}$CN absorption lines are indicated at the bottom in cyan.
            }
   \label{fig:c13}
\end{figure}

\paragraph{Light metals: Sodium to potassium.}
For the following elements, we used the most prominent absorption features in the synthetic spectra, which show the largest impact of a change of the abundance \sT{tab:abundlines}. The Na doublet \Jonww{Na}{i}{5890.8, 5896.5} \sA{fig:lm_abund} is used to find a Na abundance of $[\mathrm{Na}/\mathrm{Fe}] = \textcolor{black}{0.1\pm 0.7,}$ which resembles well with the sample of Ba stars analyzed by \citet[][\textcolor{black}{\ab{fig:abund_vgl}}]{decastroetal2016} .
From a fit to the regions affected by absorption due to Al we derive $[\mathrm{Al}/\mathrm{Fe}] = \textcolor{black}{-0.3 \pm 1.0}$. 
Ba stars typically show a slight enhancement \textcolor{black}{in Al, which is also present in the sample of \citet{decastroetal2016}}. For S we derive $[\mathrm{S}/\mathrm{Fe}] = \textcolor{black}{1.2 \pm 1.2}$ and for K we found $[\mathrm{K}/\mathrm{Fe}] = \textcolor{black}{0.1 \pm 1.0}$.

\paragraph{Iron-peak elements: Calcium to copper.}

Selected absorption features due to neutral Ca to Cu are shown in Figs.\,\ref{fig:m1_abund} and \ref{fig:m2_abund}. We performed a fit for a set of \Ion{Fe}{1} and \Ion{Fe}{2} absorption lines and found this star to be metal-poor with $[\mathrm{Fe}/\mathrm{H}] = \textcolor{black}{-0.3 \pm 1.0}$. For all iron-peak elements prior to Fe, we found solar values or slight underabundances, namely $[\mathrm{Ca}/\mathrm{Fe}] = \textcolor{black}{-0.1 \pm 1.0}$, $[\mathrm{Ti}/\mathrm{Fe}] = \textcolor{black}{-0.6 \pm 1.4}$, $[\mathrm{V}/\mathrm{Fe}] = \textcolor{black}{-0.6 \pm 1.0}$,  $[\mathrm{Cr}/\mathrm{Fe}] = \textcolor{black}{-0.2 \pm 1.3}$, and $[\mathrm{Mn}/\mathrm{Fe}] = \textcolor{black}{0.0 \pm 1.0}$. For Sc, we determined an upper limit of $[\mathrm{Sc}/\mathrm{Fe}] = \textcolor{black}{-0.4}$. These low values for the $\alpha$ elements Ca and Ti do not correspond with the trend in the sample of \citet{decastroetal2016} who found an enrichment of these elements compared to Fe with decreasing metallicity. The values close to solar for the other elements, however, are in good agreement with the sample showing a clustering of the abundances around $[\mathrm{X}/\mathrm{Fe}] = 0.0$. For the elements of this group subsequent to Fe, we determined overabundances compared to Fe of $[\mathrm{Co}/\mathrm{Fe}] = \textcolor{black}{0.5 \pm 1.0}$, $[\mathrm{Ni}/\mathrm{Fe}] = \textcolor{black}{0.6 \pm 1.3}$, and $[\mathrm{Cu}/\mathrm{Fe}] = \textcolor{black}{1.2 \pm 1.5}$.

\paragraph{Trans-iron elements: Zinc to osmium.}

Selected wavelength regions that are among those showing the largest impact of a change in abundance for the elements of this group are shown in Figs.\,\ref{fig:te1_abund}-\ref{fig:te3_abund}. Since we found only few significant absorption features for the majority of these elements, the statistical errors of the determined values are large (often $> 1$\,dex) and in many cases only upper abundance limits could be established. 
We obtained $[\mathrm{Zn}/\mathrm{Fe}] < \textcolor{black}{1.6}$ and $[\mathrm{Rb}/\mathrm{Fe}] = \textcolor{black}{1.4 \pm 1.3}$.
For the elements around the first peak of the s-process, we found $[\mathrm{Sr}/\mathrm{Fe}] = \textcolor{black}{1.0 \pm 1.5}$, $[\mathrm{Y}/\mathrm{Fe}] = \textcolor{black}{0.4 \pm 1.5}$, and $[\mathrm{Zr}/\mathrm{Fe}] = \textcolor{black}{0.2 \pm 1.5}$. Despite the large uncertainties, we find a good agreement with the Sr abundances of the sample of \citet[][\textcolor{black}{\ab{fig:abund_vgl}}]{karinkuzhietal2018}. The Sr abundances for most of the Ba stars of \citet{merleetal2016} are slightly lower but some reach up to 1\,dex as well. Also the Y abundance of \pnh agrees well with the sample of \citet{merleetal2016}, whereas the majority of Ba giants of \citet{karinkuzhietal2018} and \citet{decastroetal2016} crowd around higher values of Y. The $[\mathrm{Zr}/\mathrm{Fe}]$ abundances of \citet{karinkuzhietal2018} are significantly higher (all $ >$1\,dex), whereas some of the Ba stars of \citet{merleetal2016} agree with low values of $[\mathrm{Zr}/\mathrm{Fe}]$. Also the sample of \citet{decastroetal2016} clusters around $[\mathrm{Zr}/\mathrm{Fe}] \approx 1$ but also shows stars with comparatively low values like that for \pnh. For the elements subsequent to this peak, we determined $[\mathrm{Nb}/\mathrm{Fe}] < \textcolor{black}{1.0}$, $[\mathrm{Mo}/\mathrm{Fe}] = \textcolor{black}{1.4 \pm 1.3}$, and $[\mathrm{Ru}/\mathrm{Fe}] < \textcolor{black}{2.1}$.\\
One key element of this analysis is the radioactive Tc. The strongest absorption features that appear in the synthetic spectra are \Ionww{Tc}{1}{4031.6, 4095.7, 4238.2, 4262.3, 4297.1}. Unfortunately, these lines could not be clearly identified in the observed spectrum but it could be used to establish an upper abundance limit of $\log \epsilon_\mathrm{Tc} < 2.5$\footnote{$\log \epsilon = 12 + \log (n_\mathrm{X} / n_\mathrm{H} )$}.\\
For the determination of the Ba abundance, we used \Ionww{Ba}{2}{4554.0, 4931.1, 5853.7, 6141.7, 6496.9}. The first two are very strong and sensitive to small variations of the abundance \sA{fig:te2_abund}. This helped to constrain $[\mathrm{Ba}/\mathrm{Fe}] = \textcolor{black}{1.8 \pm 0.5,}$ which agrees within the error limits with the previous value of \citet{miszalskietal2013hen}. The values determined by \citet{karinkuzhietal2018} for this element range from $0.81 \leq [\mathrm{Ba}/\mathrm{Fe}] \leq 2.67$. Also the sample of \citet{merleetal2016} shows a scatter between almost solar and 2.5. Our strong enrichment found for \pnh, thus, is not exceptional \textcolor{black}{\sA{fig:abund_vgl}}.  \\
\Ion{La}{2} absorption lines yield $[\mathrm{La}/\mathrm{Fe}] = \textcolor{black}{1.5 \pm 1.6}$. For the other elements of the second peak of the s-process we found $[\mathrm{Ce}/\mathrm{Fe}]  < \textcolor{black}{2.3}$, $[\mathrm{Pr}/\mathrm{Fe}] < \textcolor{black}{2.7}$, and $[\mathrm{Nd}/\mathrm{Fe}] = \textcolor{black}{0.8 \pm 1.5}$. The upper limit for Ce lies above the value range of $1.02 \leq [\mathrm{Ce}/\mathrm{Fe}] \leq 1.76$ determined by \citet{karinkuzhietal2018} and also above that of \citet{merleetal2016} ranging from solar to 1.5\,dex. The Ba stars of \citet{decastroetal2016} cluster around an enrichment of 1.0\,dex and none of these stars reach values above 2.5\,dex. Our upper limit for Pr lies above the value range of $1.18 \leq [\mathrm{Pr}/\mathrm{Fe}] \leq 2.55$ of \citet{merleetal2016}. However, our value for the Nd abundance is below their range of values. The star with the lowest Nd abundance shows $[\mathrm{Nd}/\mathrm{Fe}] = 1.18$, whereas the sample of \citet{decastroetal2016} clustering around $[\mathrm{Nd}/\mathrm{Fe}] \approx 1.0$ perfectly agrees with our value within the error limits. \\
For the further rare-earth elements Sm, Eu, Gd, Tb, Dy, and Er, we could only establish upper abundance limits \sT{tab:abund}. These elements are not analyzed by \citet{merleetal2016} and \citet{decastroetal2016}. \citet{karinkuzhietal2018} found ranges of $1.02 \leq [\mathrm{Sm}/\mathrm{Fe}] \leq 2.17$, $0.96 \leq [\mathrm{Eu}/\mathrm{Fe}] \leq 1.43$, and $1.60 \leq [\mathrm{Dy}/\mathrm{Fe}] \leq 2.51$. Our upper limits of $[\mathrm{Sm}/\mathrm{Fe}] < \textcolor{black}{1.1}$, and $[\mathrm{Eu}/\mathrm{Fe}] < \textcolor{black}{1.0}$ lie within these ranges; $[\mathrm{Dy}/\mathrm{Fe}] < \textcolor{black}{3.8}$ is significantly higher. Furthermore, we could determine the abundances  of Hf, W, and Os to be $[\mathrm{Hf}/\mathrm{Fe}] < \textcolor{black}{1.4} $, $[\mathrm{W}/\mathrm{Fe}] = \textcolor{black}{1.1 \pm 1.5}$, and $[\mathrm{Os}/\mathrm{Fe}] < \textcolor{black}{1.8}$.

\section{Discussion}
\label{sect:discussion}

\subsection{Element abundances}

We compared our results with the yields from nucleosynthesis calculations of \citet{karakasetal2016} for a metallicity of $Z=0.007$, in line with the low metallicity of \textcolor{black}{$Z=0.006$ that we determined from $Z = 10^{[\mathrm{Fe}/\mathrm{H}]} Z_\odot$ with $[\mathrm{Fe}/\mathrm{H}]=-0.3$ and $Z_\odot = 0.0134$ \citep{asplundetal2009}}. From these models and those of \citet{karakasetal2018}, it becomes obvious that AGB nucleosynthesis does not affect the abundances of the iron peak elements and, thus, it seems reasonable to assume the same low metallicity for both components of the binary.\\
The fact that we cannot see a $^{13}$C enhancement agrees very well with the theoretical calculations predicting even an enhancement of the initial solar $^{12}$C/$^{13}$C-ratio for models with initial masses $1.5\,$\Msol$ \le M_\mathrm{ini} \le 4.0$\,\Msol \textcolor{black}{where our estimated initial mass \sK{sect:masstrans} lies within.} \\
The finding that the iron-peak elements prior to Fe show underabundances and those subsequent to Fe are enhanced leads to the speculation that this pattern may be caused by neutron capture on the former elements as seed species and the formation of elements heavier than Fe. \ab{fig:abund_pattern} also shows an enrichment due to AGB nucleosynthesis for the elements subsequent to Fe.\\
The observed N enhancement of $[\mathrm{N}/\mathrm{Fe}] = 0.7 \pm 0.8 $ \sA{fig:abund_pattern} is in line with the enhancement found for the Ba stars of \citet[][\ab{fig:abund_vgl}]{karinkuzhietal2018}. A high $[\mathrm{N}/\mathrm{C}]$ ratio as found for this object is discussed in the literature \citep[e.g., ][]{smiljanicetal2006,merleetal2016}. These authors argue that CN processing in Ba stars could result in higher N abundances. 
\textcolor{black}{According to \citet{smiljanicetal2006}, an increased $[\mathrm{N}/\mathrm{C}]$ ratio can be caused by mixing events such as the first DU or by a more complex mixing process due to rotation for intermediate mass stars. This would be an indicator for hydrogen burning via the CNO-cycle in the stellar core. With the assumed mass for the primary star \sK{sect:masstrans}, this should be the dominating fusion process in this star. The fast rotation of the Ba-CSPN is most likely due to transfer of angular momentum from the primary and therefore does not imply that this star was rotating exceptionally fast initially so as to affect its $[\mathrm{N}/\mathrm{C}]$ abundance ratio. 
}  

For Tc, we could not identify the presence of any line without doubt and, thus, cannot constrain the abundance further than $\log \epsilon_\mathrm{Tc} < 2.5$. Therefore, we cannot claim this star to have Tc in its atmosphere, which would directly lead to the necessity of prior mass transfer. The models of \citet{karakasetal2016} predicted a final surface abundance between $\log \epsilon_\mathrm{Tc} = 1.11$  and 1.24 for the models with initial masses between 2.1 and 2.5\,\Msol, which lies well below the upper limit for \pnh. \textcolor{black}{Another diagnostic element reflecting recent s-process nucleosynthesis is Nb. According to \citet{neyskensetal2015}, this mono-isotopic species is synthesized by the decay of the radioactive $^{93}$Zr produced by s-process nucleosynthesis. Compared to $^{99}$Tc, this species has a longer half-life time of $1.53$\,Myr. Following our estimate made for Tc \sK{sect:intro}, we do not expect an significant enrichment in Nb, since the primary's post-AGB age should be much shorter than the $^{93}$Zr half-life and, thus, a large fraction of this species should still be present. Thus, the  Nb/Zr would not represent the $^{93}$Zr/Zr ratio at the end of the AGB and cannot be employed as proof for prior mass transfer. Furthermore, the Zr abundance can be determined only within a very large error range and for Nb, we find an upper limit only.}
\\
The detection of Tc is not hampered by the resolution of the spectrograph. The limiting factor is the S/N. We estimate the needed S/N that would be necessary to clearly distinguish between a model without Tc and one with $\log \epsilon_\mathrm{Tc} = 1.2$. From \ab{fig:sntest}, it becomes clear, that the current S/N is not sufficient to determine a Tc abundance of that level. Currently, the single spectra have a S/N of 3 at that wavelength region. This is increased by co-adding all the spectra, but still a S/N increased by a factor of 3 would be necessary. According to the UVES exposure time calculator (ETC), the needed S/N would require about a six fold longer exposure. For the future Extremely Large Telescope (ELT) the estimate is more promising. By using the E-ELT Spectroscopic ETC, we find that the required S/N is reached with an exposure of about half that of a single observation used in this analysis. For stars with a lower rotational velocity, the detection would become easier \sA{fig:sntest}. Unfortunately, all Ba CSPNe that are known up to now seem to rotate fast \citep[shortest period of 4.7\,d for \object{Abell\,70} and \object{WeBo\,1} and longest period of 5.9\,d for \object{LoTr\,5}; ][]{bondetal2003,miszalskietal2012,aller18}, most likely because of the transfer of angular momentum by accretion of matter from the companion.

\begin{figure*} 
  \resizebox{\hsize}{!}{\includegraphics{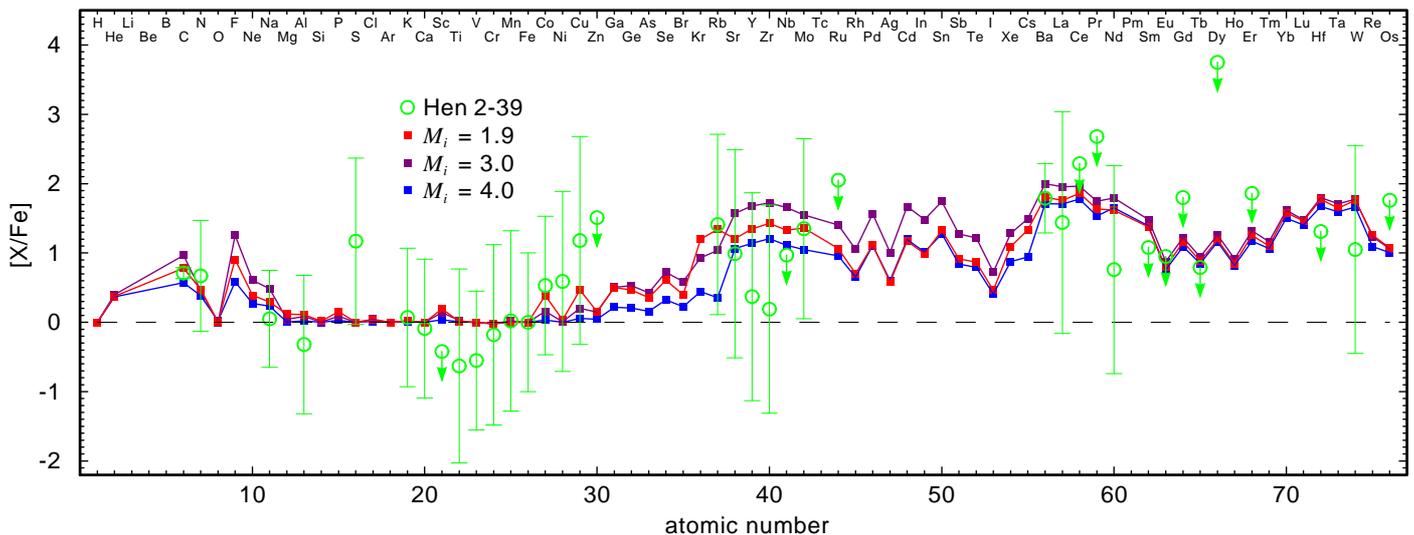}}
   \caption{Atmospheric element abundances of \pnh compared to the final yields of a selection of evolutionary models from \citet{karakasetal2016} with a metallicity of $Z=0.007$. The initial masses are indicated in the upper panel. Arrows indicate upper limits.            }
   \label{fig:abund_pattern}
\end{figure*}
\begin{figure*} 
  \resizebox{\hsize}{!}{\includegraphics{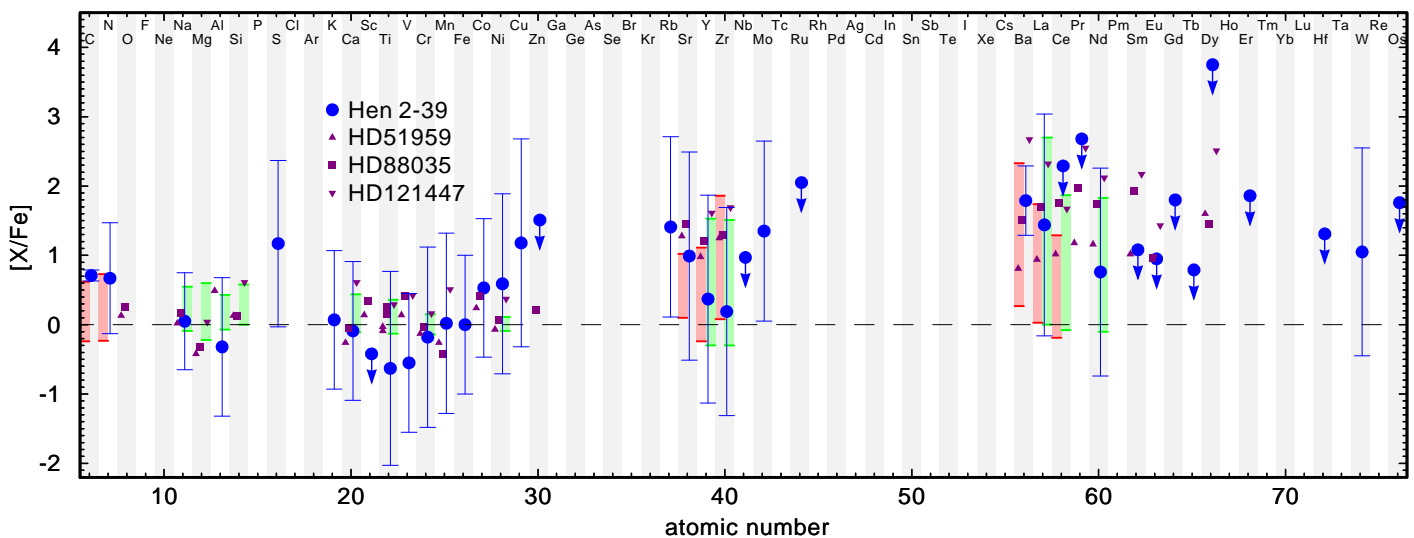}}
   \caption{Atmospheric element abundances of \pnh (blue) compared to the values for the three Ba stars \object{HD515959}, \object{HD88035}, and \object{HD121447} analyzed by \citet[][(purple)]{karinkuzhietal2018} and the ranges that \citet[][(red)]{merleetal2016} and \citet[][(green)]{decastroetal2016} found for their sets of Ba-stars. Arrows indicate upper limits.       }
   \label{fig:abund_vgl}
\end{figure*}

\begin{figure} 
  \resizebox{\hsize}{!}{\includegraphics{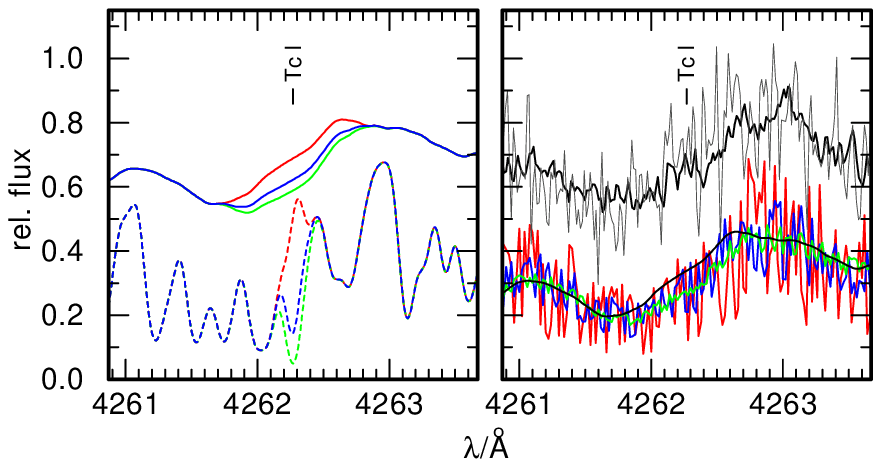}}
  \caption{Left panel: Comparison between models with $\log \epsilon_\mathrm{Tc} = 0.0, 1.2, 2.5$ (red, blue, green) for models broadened with a rotational profile ($v_\mathrm{rot} = 36$\,km/s, solid lines with 0.35 offset) and without rotation (dashed). Right: Comparison between a rotationally broadened model without Tc (black) and models with $\log \epsilon_\mathrm{Tc} = 1.2$ with different levels of artificial noise (S/N of the single observation: red, S/N of co-added spectrum: blue, required S/N: green). For comparison, the observation is shown with an offset of 0.35 (single spectrum: gray, co-added spectra: black).
            }
   \label{fig:sntest}
\end{figure}

\subsection{Mass transfer \label{sect:masstrans}}

By comparing our determined enrichment in s-process elements to the yields from evolutionary models for different initial masses of \citet{karakasetal2018}, we try to confirm that this can be the result of realistic mass transfer. 
For a primary that is currently in the stage of a CSPN, the secondary should have a mass that is lower by about 5\,\% to be currently in the evolutionary stage of a red giant (assuming a mass dependent relation for the main-sequence lifetime $t \sim M^{-2.5}$). According to \citet{jossetal1987}, these stars should have a radiative core of about 0.3\,\Msol \textcolor{black}{and a convective envelope of a mass $M_\mathrm{env}=M_\mathrm{ini} - 0.3$\,\Msol within which the accreted mass becomes diluted.} 
We want to determine a realistic mass range for the primary by comparing the total mass for the different elements that is ejected during the AGB evolution with the mass that the secondary would have needed to accrete to become that enriched. The mass of element X that needs to be accreted is given by $M_\mathrm{need} = M_\mathrm{env} (\mathrm{mf}_\mathrm{X,final} - \mathrm{mf}_\mathrm{X,initial} )$ with a final mass fraction $\mathrm{mf}_\mathrm{X,final}$ according to our analysis results and an initial mass fraction $\mathrm{mf}_\mathrm{X,initial}$ according to the low metallicity. In \ab{fig:yield_pattern}, we show the needed mass compared to the total ejected mass for C, N, and the elements heavier than Fe that show a significant production due to AGB nucleosynthesis.
Since most of the abundances could not be constrained within small error limits, we focus on the C and Ba abundances. For all models of \ab{fig:yield_pattern} and some additional models, we calculated the percentage of the total ejecta that would need to be accreted to produce the observed enrichment \sT{tab:perc}.
It is obvious that only the models for an initial mass between 1.5 and 4.0\,\Msol can explain the enrichment due to a realistic mass transfer. 
For the models with the lowest initial masses as well as for those with the highest masses, the ratio of the yield of Ba to that of C is smaller, i.e., these models produce a smaller amount of Ba compared to C.
The fraction of accreted mass of C and Ba should be equal. Thus, a {1.75-3.00\,\Msol progenitor seems to be most consistent with the abundance determinations}. \textcolor{black}{For this analysis, we used the models with the largest $^{13}$C-pocket that are available from \citet{karakasetal2016}. The larger the pocket size, the lower the yield of C and the higher that of Ba, i.e., the ratio of the yield of Ba to that of C is larger. Even for the models with the largest $^{13}$C-pocket, the percentage of the total ejecta that would need to be accreted is higher considering the Ba abundance compared to the C abundance. 
This ratio becomes worse for smaller pocket sizes. In addition to the choice of the $^{13}$C-pocket, yields of the evolutionary models are affected by uncertainties due to mass loss, convective mixing, reaction rates, and neutron poisons \citep{karakasetal2014}. These effects are not evaluated by \citet{karakasetal2016} and we take the tabulated yields without considering an error range.} {However, the progentitor mass estimate is affected by large uncertainties on the abundances and on the model yields and, thus, should be treated with caution.}
\\
This result leads to the conclusion that even such a high enrichment can reasonably be explained with realistic mass-transfer mechanisms such as wind-RLOF \citep{chenetal2017}. In this {scenario} only a small fraction of mass becomes unbound from the binary and the percentage of accreted mass ranges between 20 and 40\,\%. Simulations indicate that a binary with a wide separation, where mass transfer would act via the Bondi-Hoyle mechanism, can be ruled out since the percentage of accreted mass decreases to only 2 to 3\,\% \citep{theunsetal1996}. CE evolution would imply a short orbital period ($\leq$ a few days), which is not the case for this binary and, thus, this scenario is also ruled out.

   \begin{table}
   \centering
      \caption[]{Percentage of the total ejecta that would need to be accreted to produce the observed enrichment of \pnh for evolutionary models of \citet{karakasetal2016} with $Z= 0.007$ for different initial masses.}
         \label{tab:perc}
         \begin{tabular}{l r r}
            \hline
            \hline
            \noalign{\smallskip}
            $M_\mathrm{ini}/$\Msol  & C & Ba \\
            \noalign{\smallskip}
            \hline
            \noalign{\smallskip}
1.00 & 156 &  2215 \\ 
1.50 &  49 &    94 \\ 
1.75 &  25 &    36 \\ 
1.90 &  26 &    32 \\ 
2.10 &  14 &    17 \\ 
2.25 &  15 &    18 \\ 
2.50 &  12 &    15 \\ 
2.75 &   9 &    15 \\ 
3.00 &  10 &    15 \\ 
3.50 &  12 &    24 \\ 
4.00 &  18 &    27 \\ 
5.00 &  89 &  1311 \\ 
\noalign{\smallskip}
            \hline
         \end{tabular}
   \end{table}

\begin{figure*} 
  \resizebox{0.93\hsize}{!}{\includegraphics{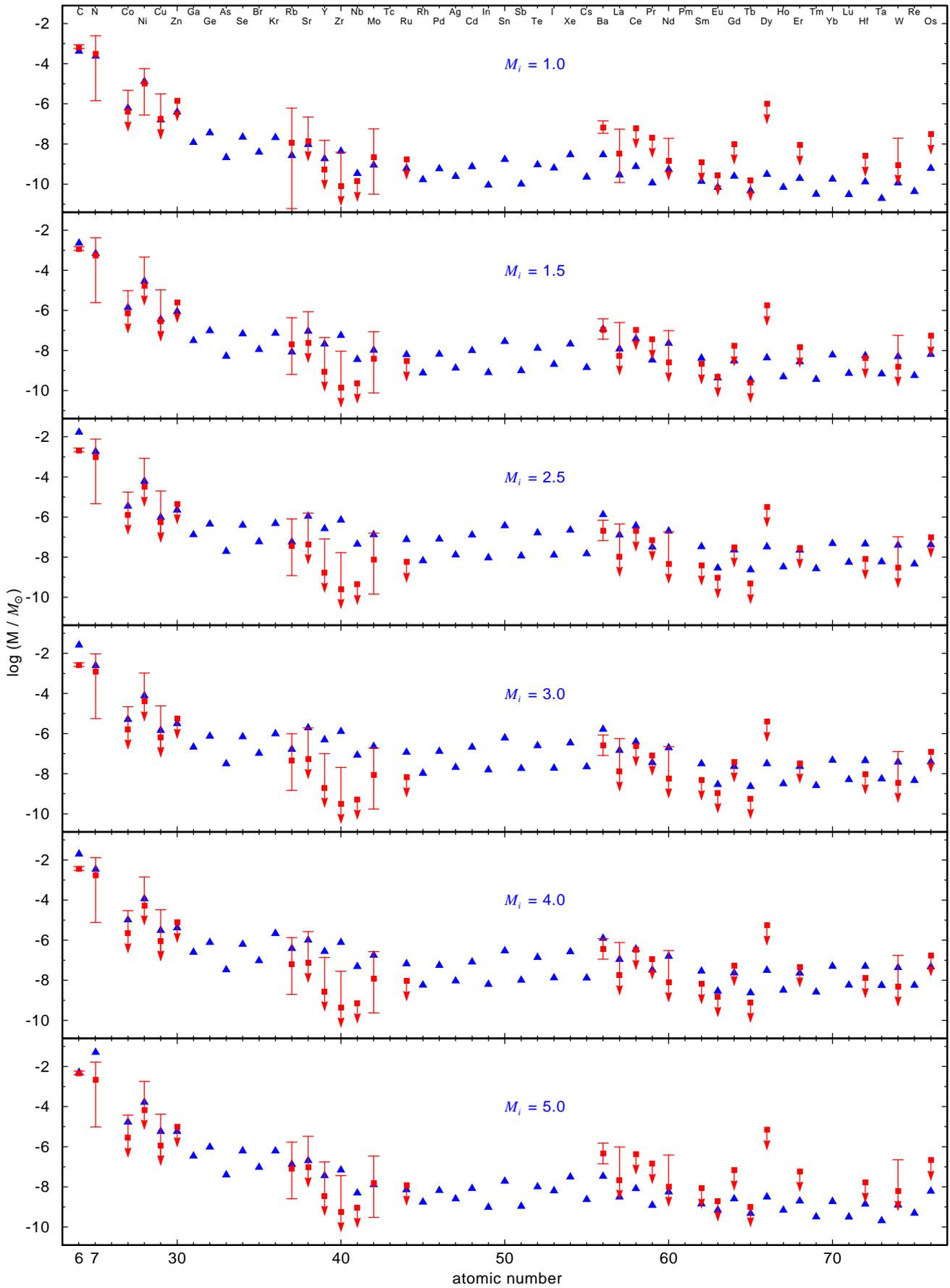}}
   \caption{Total amount of mass ejected during AGB evolution of different elements for evolutionary models from \citet{karakasetal2016} for $Z=0.007$ with initial masses of 1.0 to 5.0\,\Msol (blue, initial mass indicated in the panel) compared to the need of accreted mass to reproduce the determined abundance values of the secondary via mass transfer (red).
            }
   \label{fig:yield_pattern}
\end{figure*}

\subsection{Spectroscopic distance}
\label{ssect:spectdist}
For \pnh, several different distances are published. The nebular analysis of \citet{miszalskietal2013hen} yields 5.7\,kpc. \citet{frewetal2016} found $7.6^{+1.5}_{-1.3}$\,kpc based on the companion spectral type. The CSPN of \pnh is in the Gaia data release \citep[DR2, ID: 5256396485463285504, ][]{gaiavizier2018}. 

The parallax of $0.0564 \pm 0.0340$ mas is affected by a large error corresponding to relative errors of $60.3\,\%$. 
The object is also contained in the catalog of distances of \citet{baillerjonesetal2018} derived from the DR2 data. They found $9.071^{+2.939}_{-1.962}$\,kpc.
With our result for the companion mass, we would like to get a clue for the distance for the binary in \pnh. Using the assumed giant $\logg = 2.5$, the derived mass of about 2.2\,\Msol for the Ba star, and the determined \Teff, we get
\begin{align*}
&M_{\mathrm{bol}} = M_{\mathrm{bol,}\odot} + 2.5 \left( \log g - \log g_\odot \right) \\
&- 2.5 \log \left( \frac{M}{M_\odot}\right) - 10 \log \left( \frac{\Teff}{\Teff_\odot}\right) = 0.262.
\end{align*}
The solar values are taken from the \textit{Sun Facts Sheet} from the NASA Goddard Space Flight Center\footnote{\url{https://nssdc.gsfc.nasa.gov/planetary/factsheet/sunfact.html}, Version 29-06-2018.} and we use $M_{\mathrm{bol,}\odot} = 4.74$\,mag for the Sun derived from the standardized absolute bolometric magnitude scale defined by the international astronomical union \citep{iau2015}. The error of $M_{\mathrm{bol}}$ is dominated by the uncertainty of \logg. Assuming an uncertainty of $\Delta \logg = 0.5$ and $\Delta M = 0.5$\,\Msol, we get $\Delta M_{\mathrm{bol}} = 1.316$.
Nevertheless, we can derive the absolute visual magnitude $M_V = M_\mathrm{bol} - \mathrm{BC}(V) = 0.852 \pm 1.525$. \textcolor{black}{The bolometric correction of  $\mathrm{BC}(V) = -0.590 \pm 0.209$ is calculated using the approach of \cite{alonsoetal1999} including the preliminary values for \Teff and the Fe abundance.}  \\
We can now estimate the distance via the distance modulus but the known $V$ magnitude shows a much larger uncertainty than the more recent infrared magnitudes \sT{tab:mag}. By comparing the calculated flux normalized on the $K$ magnitude of \citet{cutrietal2003} with the $B$ and $V$ magnitudes of \citet{tylendaetal1991} and those for the $I$, $J$, and $K$ bands from \citet{epchteinetal1999}, it becomes obvious that our model agrees very well with the brightness values in all filters \sA{fig:ebv}.
Thus, we decided to rely on the precise $I$ magnitude for the distance estimation. Using the color relation for $(V-I)$ from \citet{alonsoetal1999} for the given \Teff, we find $(V-I) = 1.543^{+0.124}_{-0.110}$\,mag. This leads to an absolute $M_I$ brightness of $M_I = -0.691^{+1.635}_{-1.649}$\,mag.\\
\citet{miszalskietal2013hen} determined an extinction of $\ebv = 0.37$\,mag. With the Galactic extinction law with $R_V = 3.1$ and the relation from \citet{cardellietal1989}, we derive the total absorption for the $I$ band of $A_I = 0.5 A_V = 0.5 \times 3.1 \ebv = 0.565$\,mag. Now, we find $d = 10^{-\left( M_I - I + A_I - 5\right)/5} = 9.15^{+10.65}_{-4.90}$\,kpc. The large error is again an effect of the assumed uncertainty in \logg but, nevertheless, the value agrees with all other distance values within these limits. Furthermore, this value is very close to that derived from the GAIA parallax measurement. With a precise distance measurement, we could get a second handle on the mass of the Ba star. The mass could then be derived by comparing its properties to evolutionary tracks for different masses and compared with our value derived using the abundance yields of evolutionary models. \\
With its Galactic latitude of $-4.239$\degr \citep{gaiavizier2018}, the star is located $0.676^{+0.789}_{-0.362}$\,kpc below the Galactic plane, which means that it is just below the edge of the Galactic thin disk \citep{rixetal2013} and, thus, should belong to the thick disk. This assignation is in agreement with the observed low metallicity since simulations for the Galactic metallicity distribution predict negative metallicity gradients for low scale heights and may change the sign at about a scale height of 1.5\,kpc. \citet{ivezicetal2012} also found thick disk stars to be more metal poor (median $[\mathrm{Fe}/\mathrm{H}] = -0.6$) compared to thin disk stars (median $[\mathrm{Fe}/\mathrm{H}] = -0.2$), where \pnh lies just in the middle.
Furthermore, we can use $M_\mathrm{bol}$ to estimate the luminosity $L / L_\odot = 10^{\left( M_{\mathrm{bol,}\odot} - {M_\mathrm{bol}} \right) / 2.5} = 61.83^{+145.95}_{-43.43}$ and the radius 
\begin{align*}
\frac{R}{R_\odot} = \sqrt[]{\frac{L}{L_\odot} \frac{T_\mathrm{eff \odot} ^4}{T_\mathrm{eff}^4}}  = 13.84^{+13.38}_{-6.78} \quad .
\end{align*}
Using the rotational velocity from Section\,\ref{subsec:rotation} and the radius the star should have a rotational period of 18.4\,d for a high inclination of $i = 90\degr$. \cite{miszalskietal2013hen} detected a photometrically variability of the star with a period of 5.46\,d. Assuming this value for the rotation, we can find an inclination of $i = 17.22\degr\,^{+18.30}_{-8.53}$. The parameters are summarized in \ta{tab:starpar}. {We speculate that the rotational axis of the giant is} perpendicular to the binary orbital plane. The low inclination is then in good agreement with the ring-like appearance of the nebula \citep{miszalskietal2013hen}, which indicates a nearly pole-on view and therefore a binary orbital plane almost in the plane of the sky \citep{hillwig16}.

   \begin{table}
   \centering
      \caption[]{Brightnesses in different filters for the Ba CSPN of \pnh.}
         \label{tab:mag}
         \begin{tabular}{l l l}
            \hline
            \hline
            \noalign{\smallskip}
Filter  &  Magnitude  & Reference\\
            \noalign{\smallskip}
            \hline
            \noalign{\smallskip}
$B$ & $17.9 \pm 0.5$    & \citet{tylendaetal1991} \\
$V$ & $16.5 \pm 0.5$    & \citet{tylendaetal1991} \\
$I$ & $14.68 \pm 0.03$  & \citet{epchteinetal1999} \\
$J$ & $13.474 \pm 0.033$  & \citet{cutrietal2003} \\
$H$ & $12.614 \pm 0.033$  & \citet{cutrietal2003} \\
$K$ & $12.338 \pm 0.030$  & \citet{cutrietal2003} \\
\noalign{\smallskip}
            \hline
         \end{tabular}
   \end{table}

\begin{figure}[t]
  \resizebox{\hsize}{!}{\includegraphics{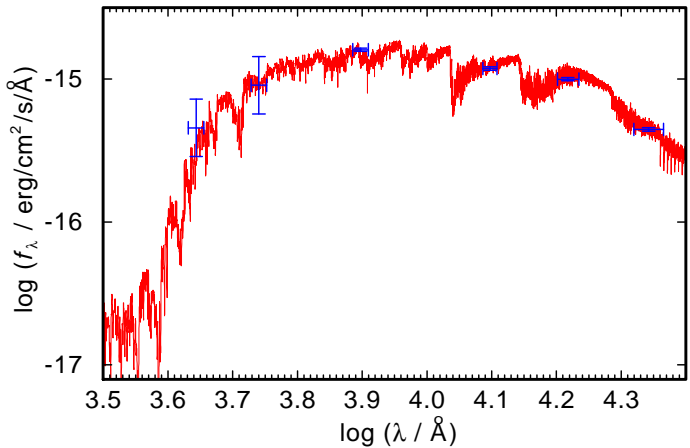}}
   \caption{Synthetic spectrum of our best model of \pnh normalized to the $H$ magnitude of \citet{cutrietal2003} and convolved with a Gaussian with $\mathrm{FWHM} = 5$\,{\AA} for clarity,
            including interstellar reddening with \ebvw{0.37} (red) compared with the observed magnitudes \sT{tab:mag}.
            }
   \label{fig:ebv}
\end{figure}

   \begin{table}
   \centering
      \caption[]{Properties of the CSPN of \pnh.}
         \label{tab:starpar}
         \begin{tabular}{l c l}
            \hline
            \hline
            \noalign{\smallskip}
\Teff            &  $(4350 \pm 150)$\,K                  \\
distance\tablefootmark{a}   &  $5.7$\,kpc     \\
distance\tablefootmark{b}   &  $\left(9.071^{+2.939}_{-1.962}\right)$\,kpc     \\
distance         &  $\left(9.15^{+10.65}_{-4.90}\right)$\,kpc   \\
BC($V$)          &  $(-0.590 \pm 0.209)$\,mag \\
$M_\mathrm{bol}$ & $0.262 \pm 1.316$\\
$(V-I)_0$         &  $1.543^{+0.124}_{-0.110}$ \\
$E(B-V)\tablefootmark{a}$         &  $-0.37$ \\
$L / L_\odot$    &  $61.83^{+145.95}_{-43.43}$ \\
$M / M_\odot$    &  $2.2 \pm 0.5$     \\
\logg            &  $2.5 \pm 0.5$    \\
$R / R_\odot$    &  $13.84^{+13.38}_{-6.78}$    \\
rotation period\tablefootmark{a}  &  5.46\,d \\
$v_\mathrm{rot} \sin i$& $38 \pm 5\,\mathrm{km}/\mathrm{s}$  \\
$i$              &  $\left(17.22^{+18.30}_{-8.53} \right)\degr$\\
            \noalign{\smallskip}
            \hline
         \end{tabular}
         \tablefoot{~\\
         \tablefoottext{a}{\citet{miszalskietal2013hen}}
         \tablefoottext{b}{\citet{baillerjonesetal2018}}
}

   \end{table}

\section{Summary and conclusions \label{sect:rescon}}

We presented and discussed the spectral analysis of UVES spectra of the Ba CSPN of \pnh.
Within the error limits, we confirm the result of \citet{miszalskietal2013hen} that the observed nucleus of \pnh has a cool atmosphere of $\Teff = (4350 \pm 150)$\,K. Furthermore, we confirm the C and Ba enrichment but can significantly improve the abundances of $[\mathrm{C}/\mathrm{H}] = 0.36 \pm \textcolor{black}{0.08} $ and $[\mathrm{Ba}/\mathrm{Fe}] = \textcolor{black}{1.8 \pm 0.50} $ due to the high resolution of the analyzed spectra.
We determined abundances or upper abundance limits for 26 trans-iron elements for the first time.
For Tc, the lightest element \textcolor{black}{with no stable isotope}, we find an upper abundance limit of $\log \epsilon_\mathrm{Tc} < 2.5$. This does not confirm the presence of Tc in the atmosphere of the star \textcolor{black}{proving} prior mass transfer. The limiting factor is not the resolution of the spectrum. For a clear detection of Tc, an exposure time about six times longer than that of all the spectra combined 
would be required to obtain the necessary S/N ratio.
We can find a low metallicity of $[\mathrm{Fe}/\mathrm{H}] = \textcolor{black}{-0.3 \pm 1.0} $ for the Ba giant.
The determined abundance pattern requires mass transfer from a companion with an extremely high enrichment of AGB nucleosynthesis products. The comparison with nucleosynthesis models of \citet{karakasetal2016} indicates an initial mass of {1.75-3.00\,\Msol} for the primary.
The percentage of ejected mass that needs to be accreted indicates that the preferred mass transfer mechanism is wind-RLOF. A wide binary involving Bondi-Hoyle accretion can be ruled out as can a CE evolution.
For this star, the distance is rather uncertain. Thus, it cannot be used for a spectroscopic determination of the mass by interpolation from evolutionary tracks. A precise spectroscopic determination of the distance is hampered by the fact that \logg cannot be constrained within narrow error limits from the analysis of the spectra. It is highly desirable to get a more precise distance measurement. This would also help to get a second value for the mass of the Ba CSPN to compare with that derived from the comparison with AGB models. {With this second measurement, it would be possible to refine the primary mass estimate and place stronger constraints on the mass transfer.} 
Our result for the height above the Galactic plane places this system among the thick disk population,  in good agreement with the subsolar metallicity derived by our analysis.
\begin{acknowledgements}
    We thank the anonymous referee for their constructive review of the manuscript.
      We thank Brent Miszalski and Thomas Rauch for their helpful comments and suggestions.
      LL is supported by the German Research Foundation (DFG, grant WE\,1312/49-1) and by the Studentship Programme of the European Southern Observatory.  DJ gratefully acknowledges the Spanish Ministry of Economy and Competitiveness (MINECO) under the grant AYA2017-83383-P.
      This research has made use of 
NASA's Astrophysics Data System and
the SIMBAD database, operated at CDS, Strasbourg, France.
This work has made use of data from the European Space Agency (ESA) mission
{\it Gaia} (\url{https://www.cosmos.esa.int/gaia}), processed by the {\it Gaia}
Data Processing and Analysis Consortium (DPAC,
\url{https://www.cosmos.esa.int/web/gaia/dpac/consortium}). Funding for the DPAC
has been provided by national institutions, in particular the institutions
participating in the {\it Gaia} Multilateral Agreement.

\end{acknowledgements}


\bibliographystyle{aa}
\bibliography{hen2-39}

\appendix

\section{Additional figures and tables.}
\label{app:additional}

\begin{figure*} 
  \resizebox{\hsize}{!}{\includegraphics{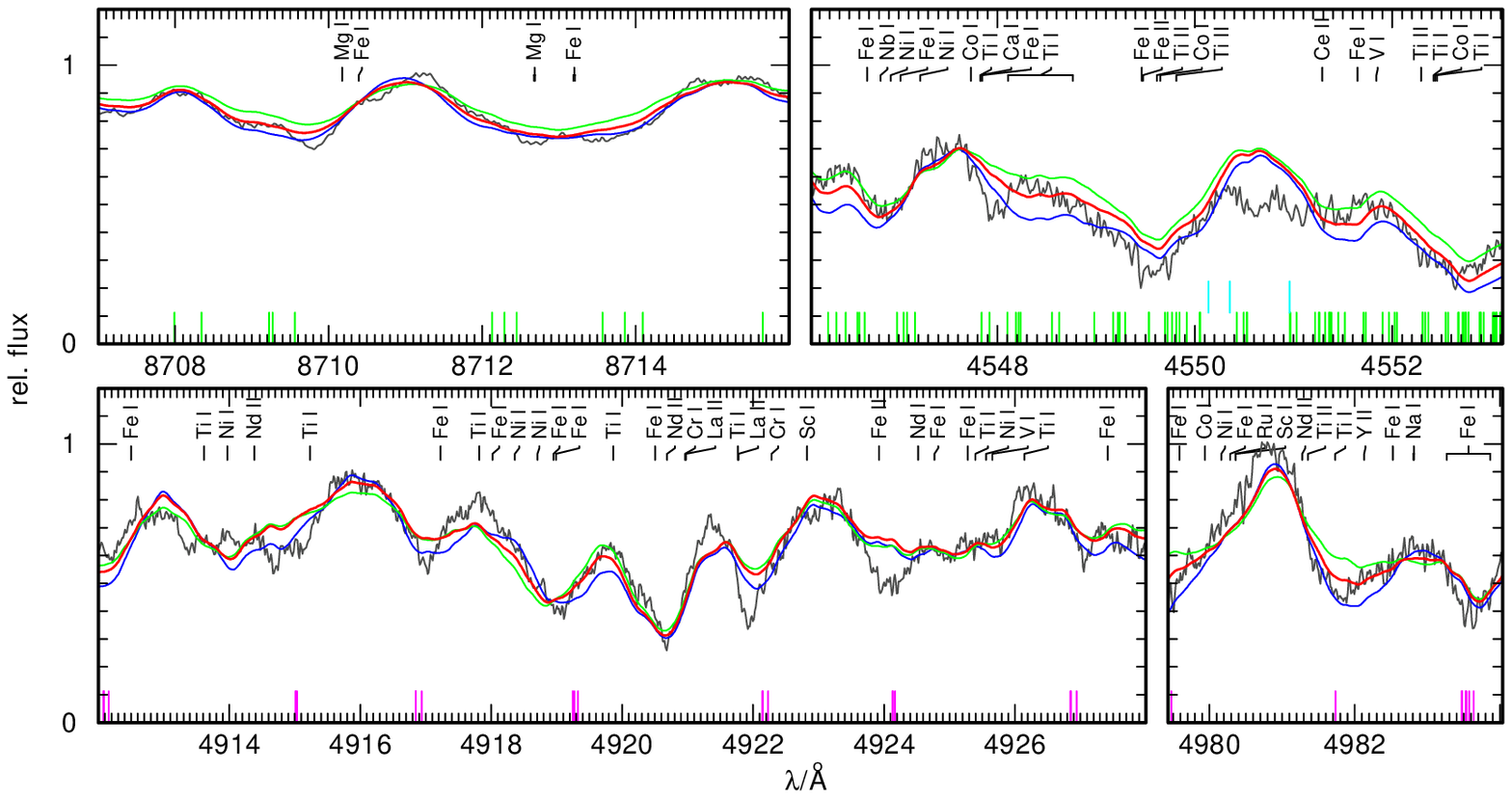}}
   \caption{Observation (gray) of \pnh compared to model spectra with $\Teff = 4000, 4250, 4500$\,K (blue, red, green) for selected regions that were used for the determination of \Teff. CN, C$_2$, and CH absorption lines are indicated at the bottom in green, purple, and cyan, respectively. All absorption lines that appear with an equivalent width $\ge 20$\,m{\AA} in the calculated spectrum are indicated.
            }
   \label{fig:teff}
\end{figure*}
\begin{figure*} 
  \resizebox{\hsize}{!}{\includegraphics{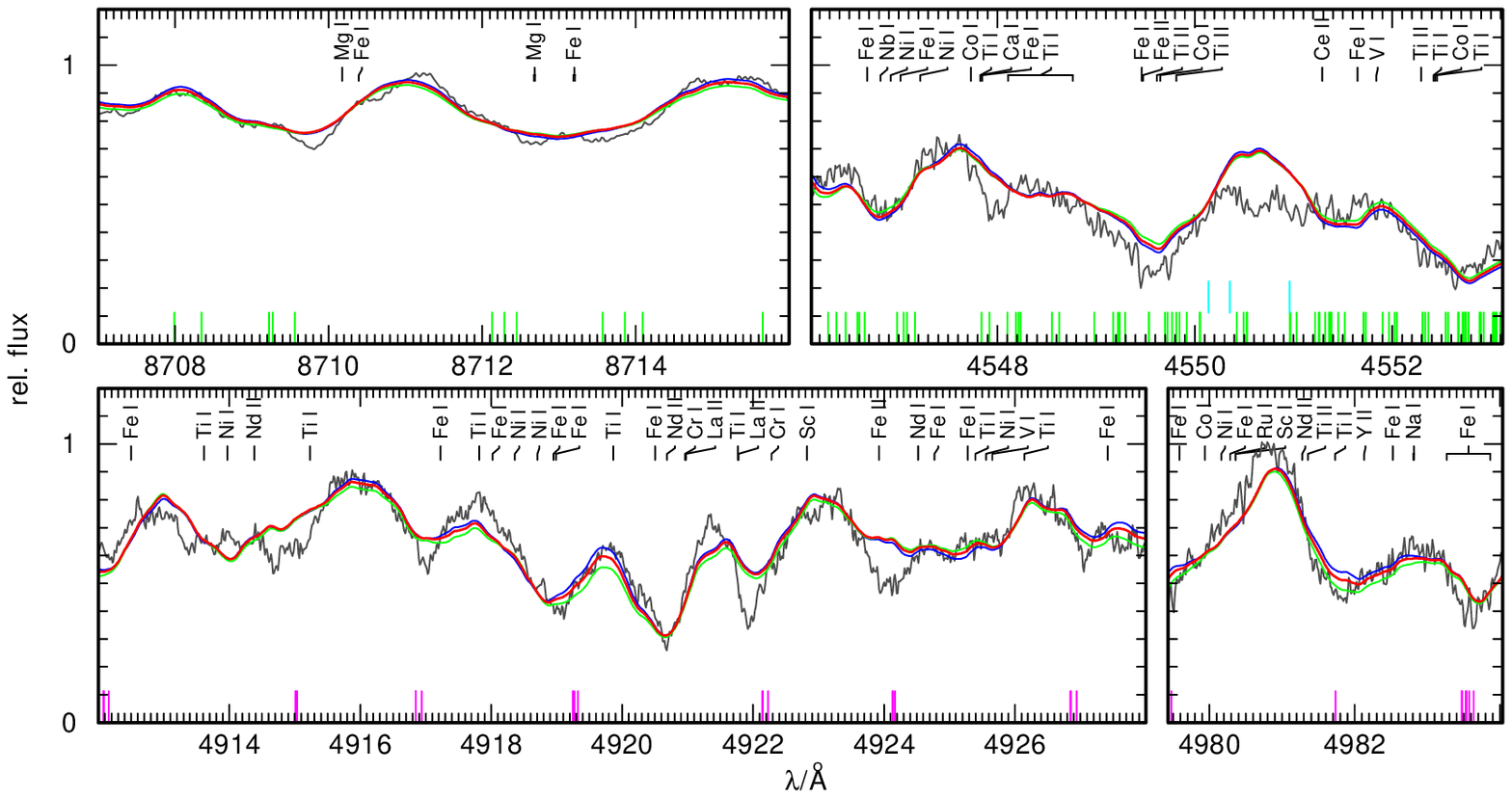}}
   \caption{Like \ab{fig:teff}, for $\logg = 2.0, 2.5, 3.0$ (blue, red, green).
            }
   \label{fig:logg}
\end{figure*}

   \begin{table}
   \centering
      \caption[]{Observation log for the UVES observations.}
         \label{tab:obslog}
         \begin{tabular}{l c c r}
            \hline
            \hline
            \noalign{\smallskip}
Start Time (UT)  &  $\lambda\,/\,{\AA}$  &  Exp. Time\,/\,s  & \textcolor{black}{S/N}\\
            \noalign{\smallskip}
            \hline
            \noalign{\smallskip}
2014-04-03 02:42:47 & 5654$-$9465 & 1500 & \textcolor{black}{14.8} \\
2014-04-03 02:42:51 & 3732$-$5000 & 1500 & \textcolor{black}{3.7} \\
2014-04-03 03:08:39 & 5654$-$9465 & 1500 & \textcolor{black}{14.1} \\
2014-04-03 03:08:39 & 3732$-$5000 & 1500 & \textcolor{black}{3.3} \\
2014-04-03 03:36:22 & 5654$-$9465 & 1500 & \textcolor{black}{12.6} \\
2014-04-03 03:36:26 & 3732$-$5000 & 1500 & \textcolor{black}{3.1} \\
2014-04-03 04:02:13 & 5654$-$9465 & 1500 & \textcolor{black}{14.5} \\
2014-04-03 04:02:14 & 3732$-$5000 & 1500 & \textcolor{black}{3.7} \\
2014-04-04 00:40:42 & 5654$-$9465 & 1500 & \textcolor{black}{13.5} \\
2014-04-04 00:40:46 & 3732$-$5000 & 1500 & \textcolor{black}{3.1} \\
2014-04-04 01:06:34 & 5654$-$9465 & 1500 & \textcolor{black}{13.1} \\
2014-04-04 01:06:34 & 3732$-$5000 & 1500 & \textcolor{black}{2.9} \\
2014-04-04 01:34:52 & 5654$-$9465 & 1500 & \textcolor{black}{14.8} \\
2014-04-04 01:34:59 & 3732$-$5000 & 1500 & \textcolor{black}{3.7} \\
2014-04-04 02:00:47 & 3732$-$5000 & 1500 & \textcolor{black}{3.5} \\
2014-04-04 02:00:47 & 5654$-$9465 & 1500 & \textcolor{black}{14.8} \\
2014-04-04 02:32:05 & 5654$-$9465 & 1500 & \textcolor{black}{14.5} \\
2014-04-04 02:32:11 & 3732$-$5000 & 1500 & \textcolor{black}{3.2} \\
2014-04-04 02:57:59 & 3732$-$5000 & 1500 & \textcolor{black}{3.3} \\
2014-04-04 02:58:00 & 5654$-$9465 & 1500 & \textcolor{black}{14.1} \\
2014-04-13 02:30:36 & 5654$-$9465 & 1450 & \textcolor{black}{12.9} \\
2014-04-13 02:30:40 & 3732$-$5000 & 1450 & \textcolor{black}{2.9} \\
2014-04-13 02:55:38 & 3732$-$5000 & 1450 & \textcolor{black}{3.2} \\
2014-04-13 02:55:39 & 5654$-$9465 & 1450 & \textcolor{black}{13.7} \\
2014-04-15 00:20:10 & 5654$-$9465 & 1450 & \textcolor{black}{14.3} \\
2014-04-15 00:20:14 & 3732$-$5000 & 1450 & \textcolor{black}{3.0} \\
2014-04-15 00:45:11 & 5654$-$9465 & 1450 & \textcolor{black}{11.3} \\
2014-04-15 00:45:12 & 3732$-$5000 & 1450 & \textcolor{black}{2.4} \\
2014-04-15 01:10:37 & 5654$-$9465 & 1450 & \textcolor{black}{12.9} \\
2014-04-15 01:10:41 & 3732$-$5000 & 1450 & \textcolor{black}{2.7} \\
2014-04-15 01:35:39 & 3732$-$5000 & 1450 & \textcolor{black}{2.9} \\
2014-04-15 01:35:39 & 5654$-$9465 & 1450 & \textcolor{black}{13.4} \\
2014-04-15 02:01:20 & 5654$-$9465 & 1500 & \textcolor{black}{13.1} \\
2014-04-15 02:01:24 & 3732$-$5000 & 1500 & \textcolor{black}{2.8} \\
2014-04-15 02:27:11 & 5654$-$9465 & 1500 & \textcolor{black}{13.1} \\
2014-04-15 02:27:12 & 3732$-$5000 & 1500 & \textcolor{black}{2.6} \\
            \noalign{\smallskip}
            \hline
         \end{tabular}
   \end{table}

\begin{figure} 
 \centering

 \resizebox{\hsize}{!}{\includegraphics[angle=-90]{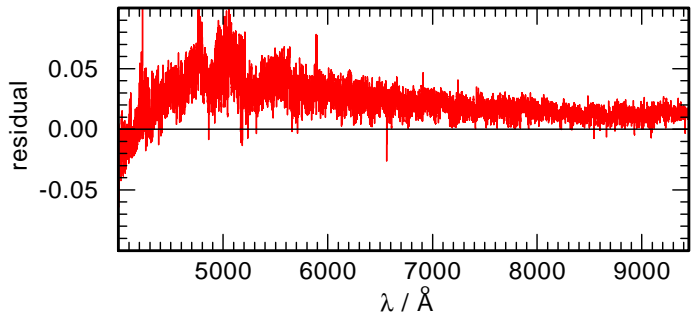}}
   \caption{Difference between the newly computed C-rich test model (model 2) and the O-rich model from the available grid (model 1) for $\Teff = 4250$\,K and $\logg = 2.5$. }
   \label{fig:res}
\end{figure}

\begin{figure*} 
 \centering

 \resizebox{0.98\hsize}{!}{\includegraphics{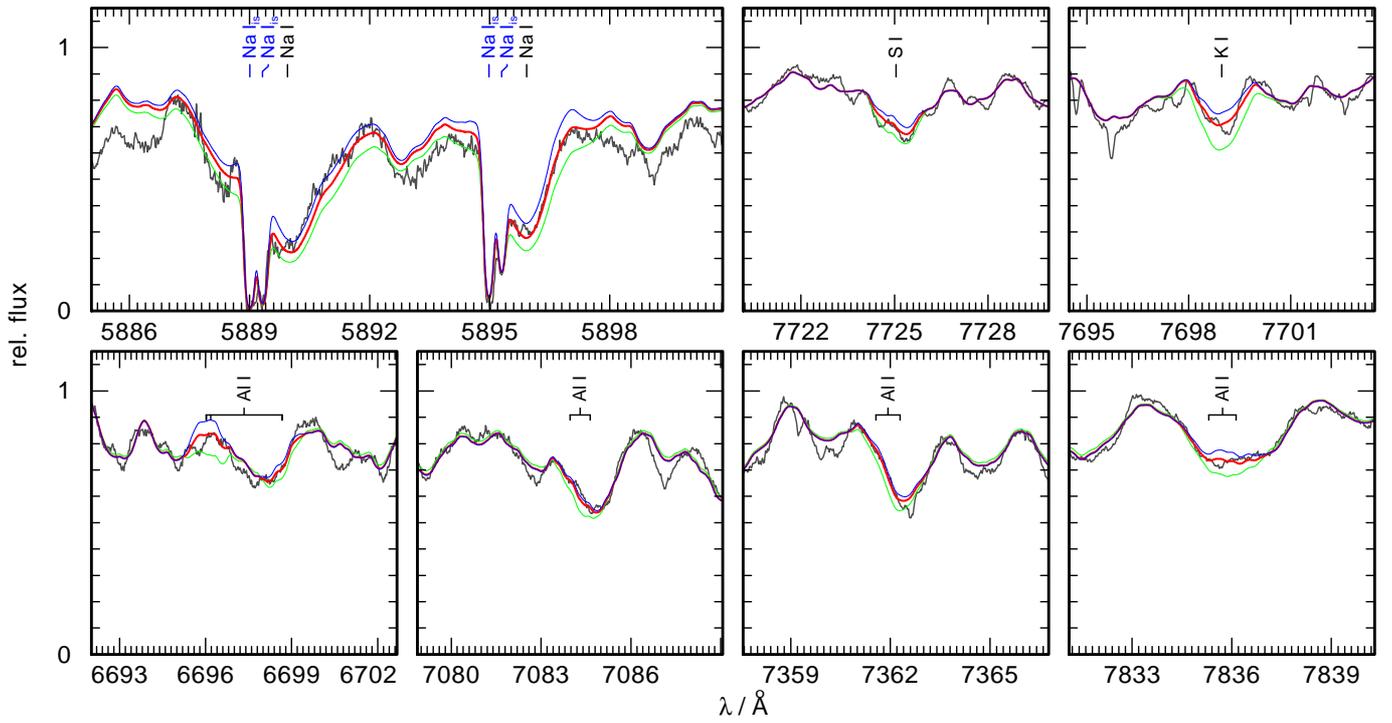}}
   \caption{Observation (gray) of \pnh compared to model spectra for selected regions around absorption lines of Na\,\textsc{i} for $[\mathrm{Na}/\mathrm{Fe}] = 0.35, 0.05, -0.25$ (green, red, and blue, respectively), S\,\textsc{i} for $[\mathrm{S}/\mathrm{Fe}] = 2.17, 1.17, 0.17$, Al\,\textsc{i} for $[\mathrm{Al}/\mathrm{Fe}] = 0.68, -0.32, -1.32$, and K\,\textsc{i} for $[\mathrm{K}/\mathrm{Fe}] = 1.07, 0.07, -0.93 $. Interstellar absorption lines are indicated with blue marks.
            }
   \label{fig:lm_abund}
\end{figure*}

\begin{figure*} 
\centering

  \resizebox{0.98\hsize}{!}{\includegraphics{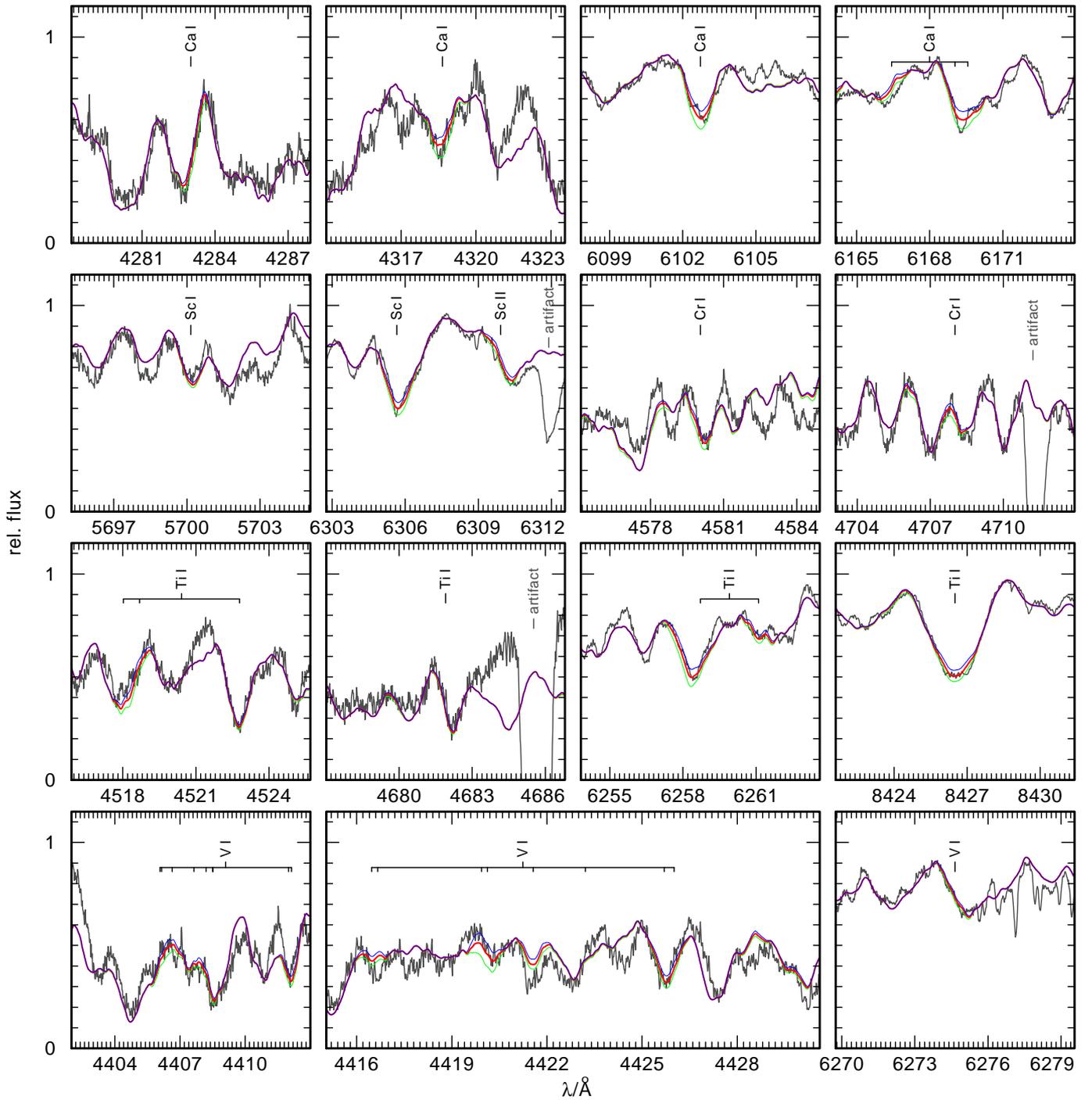}}
   \caption{Observation (gray) of \pnh compared to model spectra for selected regions around absorption lines of Ca\,\textsc{i} for $[\mathrm{Ca}/\mathrm{Fe}] = 0.41, -0.09, -0.59$ (green, red, and blue, respectively), Sc\,\textsc{i} for $[\mathrm{Sc}/\mathrm{Fe}] = 0.08, -0.42, -0.92$, Cr\,\textsc{i} for $[\mathrm{Cr}/\mathrm{Fe}] = 0.32, -0.18, -0.68$, Ti\,\textsc{i} for $[\mathrm{Ti}/\mathrm{Fe}] = -0.13, -0.63, -1.13$, and V\,\textsc{i} for $[\mathrm{V}/\mathrm{Fe}] = -0.05, -0.55, -1.05$. Artifacts arising from the overcorrection of nebula lines are indicated.
            }
   \label{fig:m1_abund}
\end{figure*}
\begin{figure*} 
 \centering

 \resizebox{0.98\hsize}{!}{\includegraphics{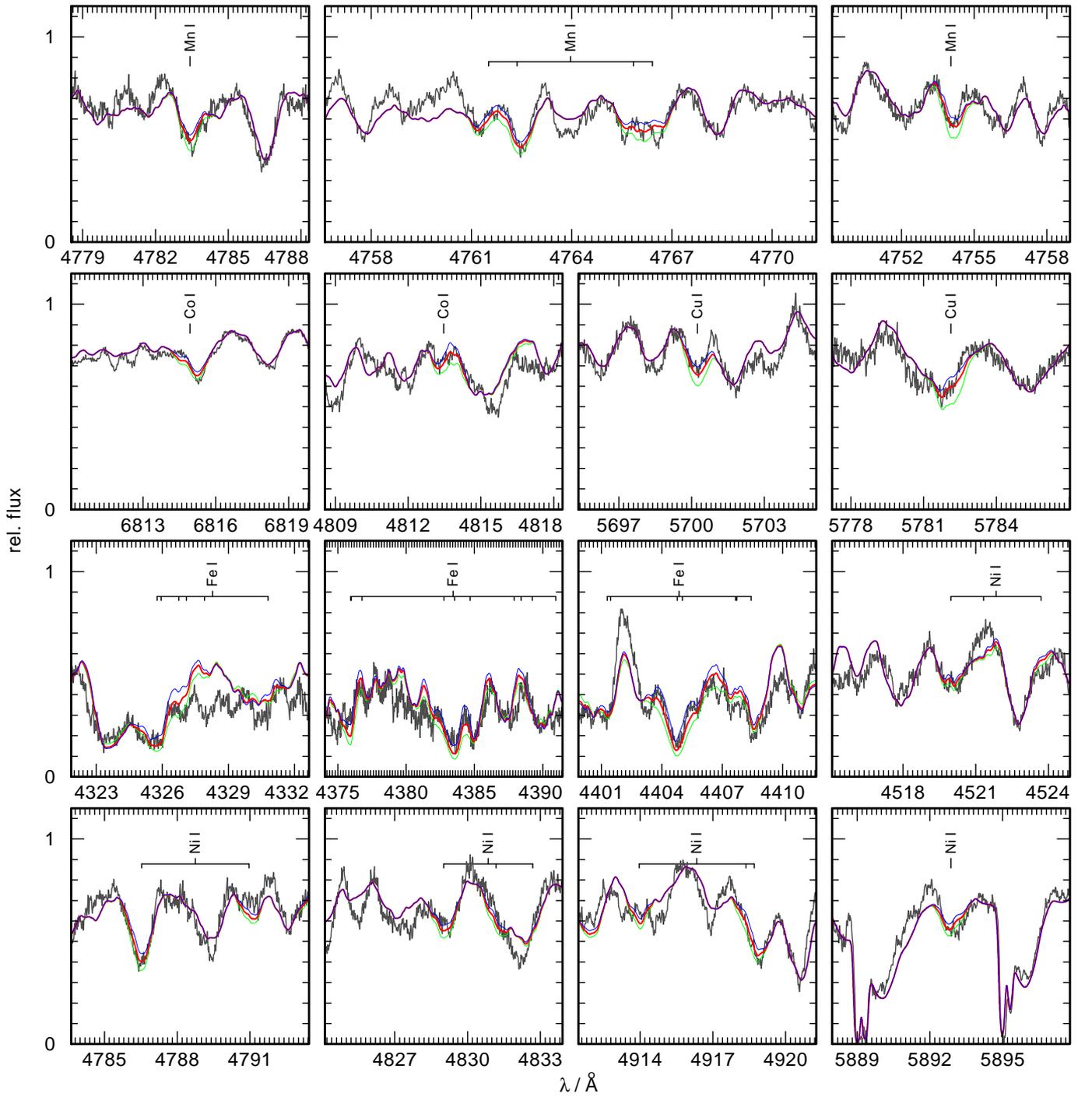}}
   \caption{Observation (gray) of \pnh compared to model spectra for selected regions around absorption lines of Mn\,\textsc{i} for $[\mathrm{Mn}/\mathrm{Fe}] = 0.52, 0.02, -0.48$ (green, red, and blue, respectively), Co\,\textsc{i} for $[\mathrm{Co}/\mathrm{Fe}] = 1.03, 0.53, 0.03$, Cu\,\textsc{i} for $[\mathrm{Cu}/\mathrm{Fe}] = 2.18, 1.18, 0.18$, Fe\,\textsc{i} for $[\mathrm{Fe}/\mathrm{H}] = 0.15, -0.35, -0.85$, and Ni\,\textsc{i} for $[\mathrm{Ni}/\mathrm{Fe}] = 1.09, 0.59, 0.09$. Interstellar absorption lines are indicated with blue marks.
            }
   \label{fig:m2_abund}
\end{figure*}

\begin{figure*} 
 \centering

 \resizebox{0.98\hsize}{!}{\includegraphics{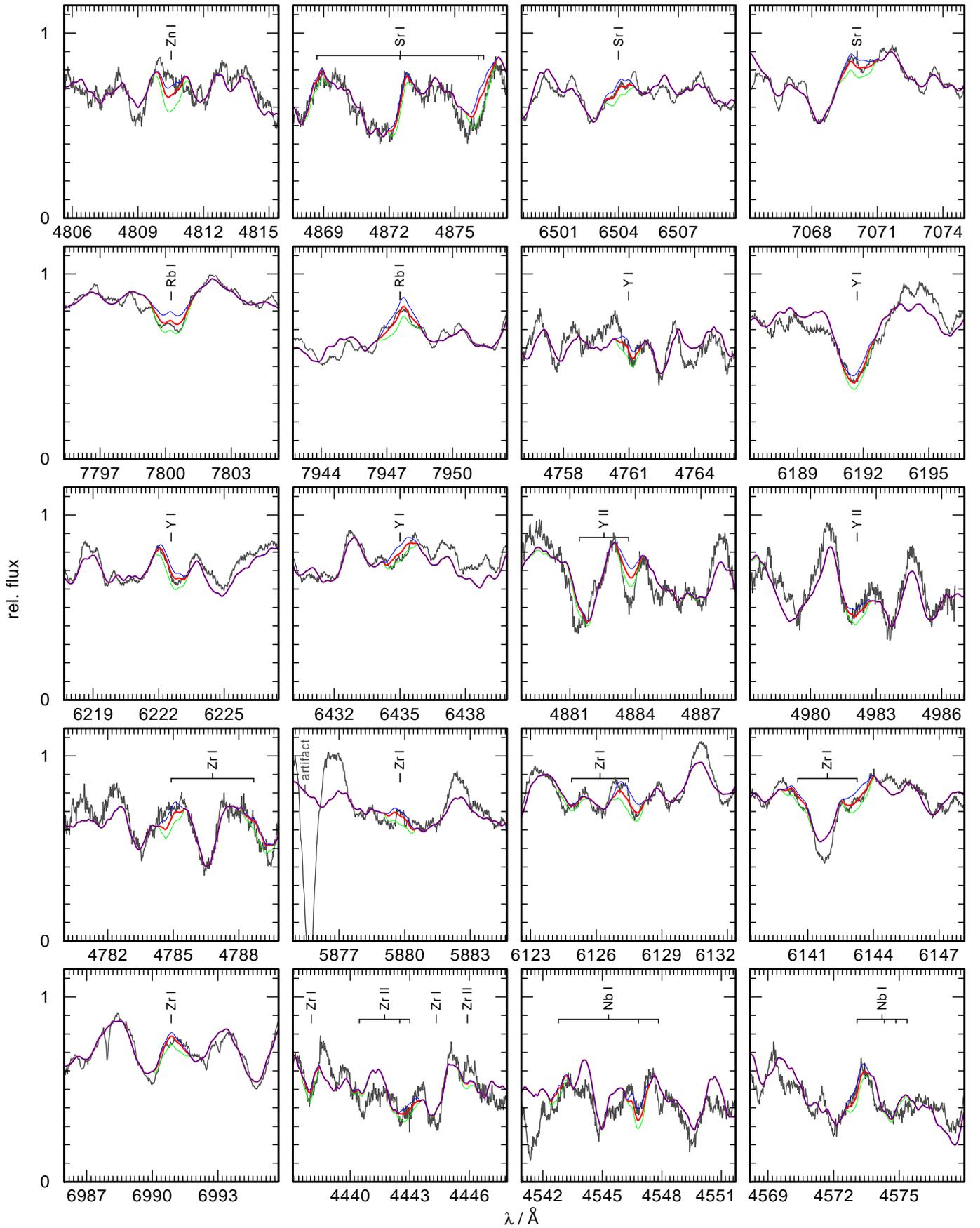}}
   \caption{Observation (gray) of \pnh compared to model spectra for selected regions around absorption lines of Zn\,\textsc{i} for $[\mathrm{Zn}/\mathrm{Fe}] = 2.51, 1.51, 0.51$ (green, red, and blue, respectively), Sr\,\textsc{i} for $[\mathrm{Sr}/\mathrm{Fe}] = 1.99, 0.99, -0.01$, Rb\,\textsc{i} for $[\mathrm{Rb}/\mathrm{Fe}] = 2.41, 1.41, 0.41$, Y\,\textsc{i} and Y\,\textsc{ii} for $[\mathrm{Y}/\mathrm{Fe}] = 1.37, 0.37, -0.63$, Zr\,\textsc{i} and Zr\,\textsc{ii} for $[\mathrm{Zr}/\mathrm{Fe}] = 1.19, 0.19 -0.81$, and Nb\,\textsc{i} for $[\mathrm{Nb}/\mathrm{Fe}] = 1.97, 0.97, -0.03$. Artifacts arising from the overcorrection of nebula lines are indicated.
            }
   \label{fig:te1_abund}
\end{figure*}

\begin{figure*}
\centering
  \resizebox{0.98\hsize}{!}{\includegraphics{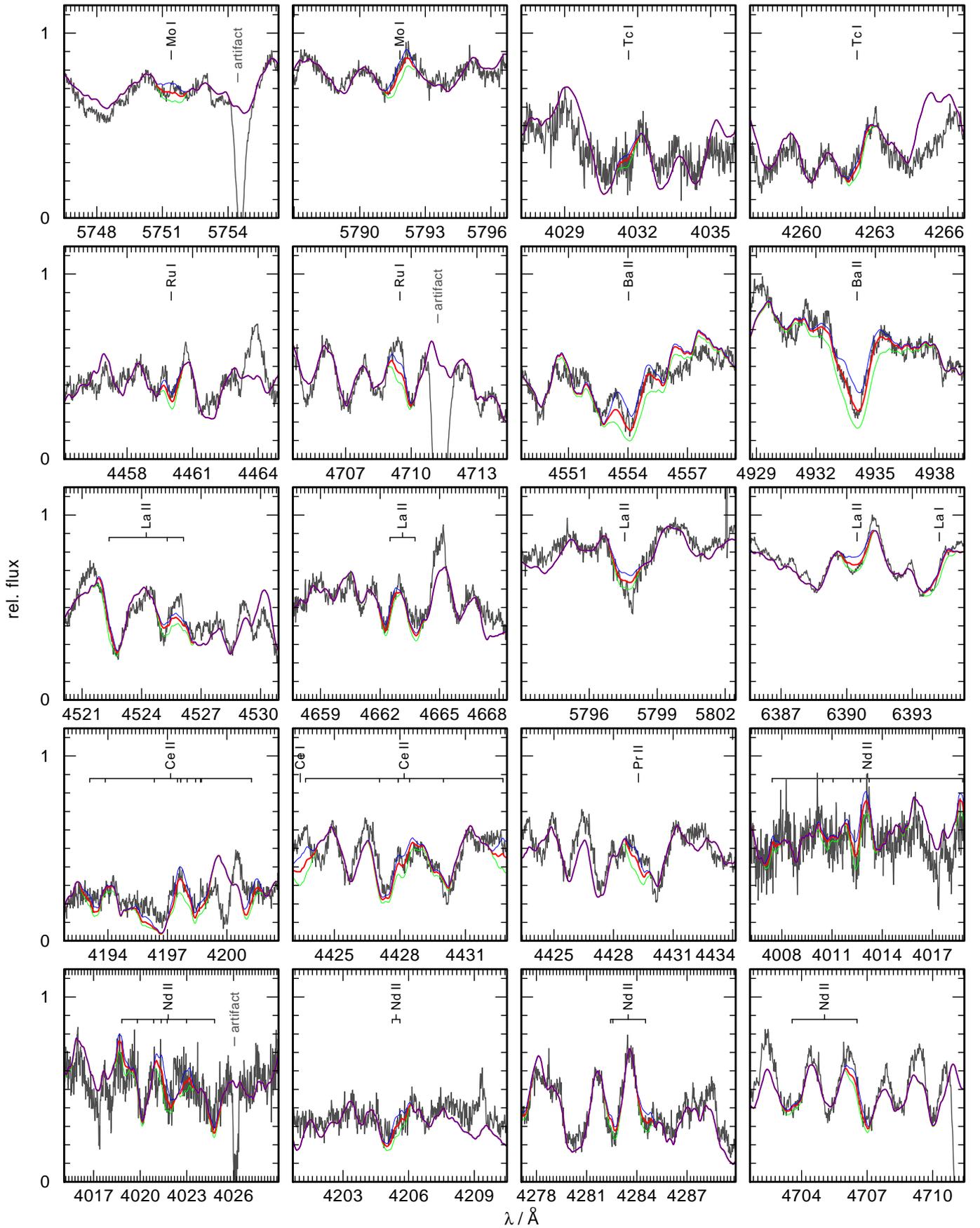}}
   \caption{Observation (gray) of \pnh compared to model spectra for selected regions around absorption lines of Mo\,\textsc{i} for $[\mathrm{Mo}/\mathrm{Fe}] = 2.35, 1.35, 0.35$ (green, red, and blue, respectively), Tc\,\textsc{i} for $\log \epsilon_\mathrm{Tc} = 3.5, 2.5, 1.5$, Ru\,\textsc{i} for $[\mathrm{Ru}/\mathrm{Fe}] = 3.05, 2.05, 1.05$, Ba\,\textsc{ii} $[\mathrm{Ba}/\mathrm{Fe}] = 2.29, 1.79, 1.27$, La\,\textsc{ii} for $[\mathrm{La}/\mathrm{Fe}] = 2.44, 1.44, 0.44$, Ce\,\textsc{i} and Ce\,\textsc{ii} for $[\mathrm{Ce}/\mathrm{Fe}] = 3.29, 2.29, 1.29$, Pr\,\textsc{ii} for $[\mathrm{Pr}/\mathrm{Fe}] = 3.68, 2.68, 1.68$, and Nd\,\textsc{i} for $[\mathrm{Nd}/\mathrm{Fe}] = 1.76, 0.76, -0.24$. Artifacts arising from the overcorrection of nebula lines are indicated.
            }
   \label{fig:te2_abund}
\end{figure*}

\begin{figure*} 
 \centering

 \resizebox{0.98\hsize}{!}{\includegraphics{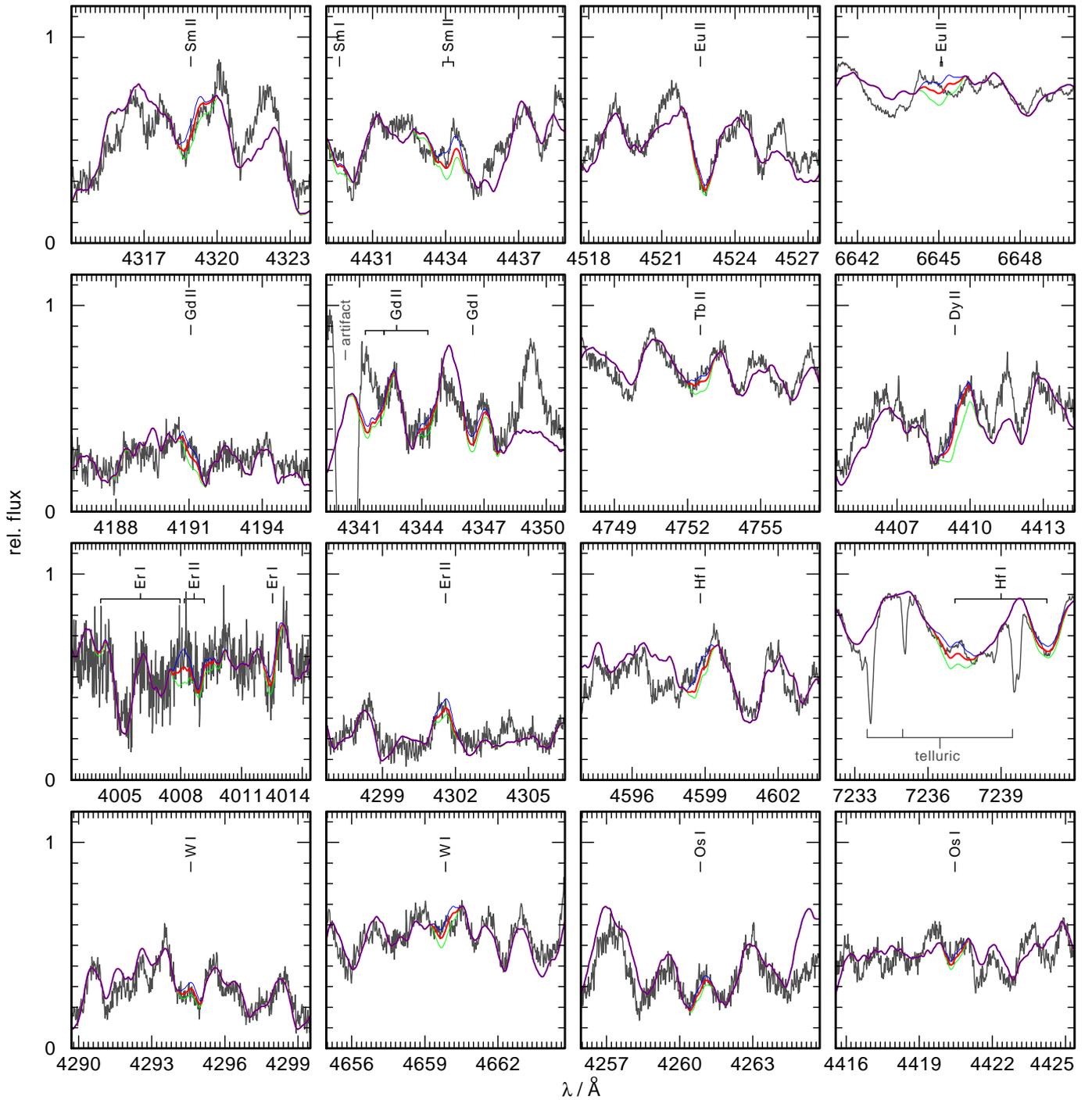}}
   \caption{Observation (gray) of \pnh compared to model spectra for selected regions around absorption lines of Sm\,\textsc{i} and Sm\,\textsc{ii} for $[\mathrm{Sm}/\mathrm{Fe}] = 2.08, 1.08, 0.08$ (green, red, and blue, respectively), Eu\,\textsc{ii} for $[\mathrm{Eu}/\mathrm{Fe}] = 1.95, 0.95, -0.05$, Gd\,\textsc{i} and Gd\,\textsc{ii} $[\mathrm{Gd}/\mathrm{Fe}] = 2.80, 1.80, 0.80$, Tb\,\textsc{ii} for $[\mathrm{Tb}/\mathrm{Fe}] = 1.79, 0.79, -0.21$, Dy\,\textsc{ii} for $[\mathrm{Dy}/\mathrm{Fe}] = 4.75, 3.75, 2.75$, Er\,\textsc{i} and Er\,\textsc{ii} $[\mathrm{Er}/\mathrm{Fe}] = 2.86, 1.86, 0.86$, Hf\,\textsc{i} for $[\mathrm{Hf}/\mathrm{Fe}] = 2.31, 1.31, 0.31$, W\,\textsc{i} for $[\mathrm{W}/\mathrm{Fe}] = 2.05, 1.05, 0.05$, and Os\,\textsc{i} for $[\mathrm{Os}/\mathrm{Fe}] = 2.76, 1.76, 0.76$. Artifacts arising from the overcorrection of nebula lines are indicated.
            }
   \label{fig:te3_abund}
\end{figure*}

\onecolumn

\begin{longtable}{c c c r r r}
\caption{\label{tab:charlines}Diagnostic lines used for the determination of stellar parameters. If no interval is given, the line belongs to the previous interval.}\\
\hline
\hline
\noalign{\smallskip}
$\lambda$ interval\,/\,{\AA}   &  $\lambda_\mathrm{line}\,/\,{\AA}$  &  Ion  &  $E_\mathrm{low}\,/\,{\mathrm{cm}^{-1}}$ & $E_\mathrm{up}\,/\,{\mathrm{cm}^{-1}}$ & $\log gf$\\
\noalign{\smallskip}
\hline
\noalign{\smallskip}
\endfirsthead
\caption{Continued.}\\
\hline\hline
\noalign{\smallskip}
$\lambda$ interval\,/\,{\AA}   &  $\lambda_\mathrm{line}\,/\,{\AA}$  &  Ion  &  $E_\mathrm{low}\,/\,{\mathrm{cm}^{-1}}$ & $E_\mathrm{up}\,/\,{\mathrm{cm}^{-1}}$ & $\log gf$\\
\noalign{\smallskip}
\hline
\endhead
\noalign{\smallskip}
\hline
\endfoot
$4020.400 \pm 1.30$  &  4020.400  &  Sc\,\textsc{i} &         0 &  24866 &  $-$0.130 \\ 
$4023.690 \pm 1.30$  &  4023.690  &  Sc\,\textsc{i} &       168 &  25014 &   0.210 \\  
$4045.820 \pm 1.30$  &  4045.820  &  Fe\,\textsc{i} &     11976 &  36686 &   0.280 \\  
$4054.544 \pm 1.30$  &  4054.544  &  Sc\,\textsc{i} &         0 &  24657 &  $-$0.750 \\  
$4063.605 \pm 1.30$  &  4063.605  &  Fe\,\textsc{i} &     12561 &  37163 &   0.072 \\  
$4071.740 \pm 1.30$  &  4071.740  &  Fe\,\textsc{i} &     12969 &  37521 &  $-$0.022 \\  
$4082.390 \pm 1.30$  &  4082.390  &  Sc\,\textsc{i} &       168 &  24657 &  $-$0.444 \\  
$4233.170 \pm 1.30$  &  4233.170  &  Fe\,\textsc{ii} &    20831 &  44447 &  $-$1.995 \\   
$4271.760 \pm 1.30$  &  4271.760  &  Fe\,\textsc{i} &     11976 &  35379 &  $-$0.164 \\  
$4305.820 \pm 1.40$  &  4305.720  &  Sc\,\textsc{ii} &     4803 &  28021 &  $-$1.200 \\   
                     &  4305.910  &  Ti\,\textsc{i} &      6843 &  30060 &   0.300 \\  
$4307.900 \pm 1.30$  &  4307.900  &  Fe\,\textsc{i} &     12561 &  35768 &  $-$0.300 \\  
$4314.080 \pm 1.30$  &  4314.080  &  Sc\,\textsc{ii} &     4988 &  28161 &  $-$0.220 \\   
$4320.750 \pm 1.30$  &  4320.750  &  Sc\,\textsc{ii} &     4883 &  28021 &  $-$0.100 \\   
$4325.010 \pm 1.30$  &  4325.010  &  Sc\,\textsc{ii} &     4803 &  27918 &  $-$0.250 \\   
$4325.760 \pm 1.30$  &  4325.760  &  Fe\,\textsc{i} &     12969 &  36079 &  $-$0.300 \\  
$4351.769 \pm 1.30$  &  4351.769  &  Fe\,\textsc{ii} &    21812 &  44785 &  $-$2.100 \\   
$4374.472 \pm 1.30$  &  4374.472  &  Sc\,\textsc{ii} &     4988 &  27841 &  $-$0.640 \\   
$4383.550 \pm 1.30$  &  4383.550  &  Fe\,\textsc{i} &     11976 &  34782 &   0.200 \\  
$4395.040 \pm 1.30$  &  4395.040  &  Ti\,\textsc{ii} &     8744 &  31491 &  $-$0.660 \\   
$4400.398 \pm 1.30$  &  4400.398  &  Sc\,\textsc{ii} &     4883 &  27602 &  $-$0.480 \\   
$4404.761 \pm 1.30$  &  4404.761  &  Fe\,\textsc{i} &     12561 &  35257 &  $-$0.142 \\  
$4415.560 \pm 1.30$  &  4415.560  &  Sc\,\textsc{ii} &     4803 &  27444 &  $-$0.510 \\   
$4443.812 \pm 1.30$  &  4443.812  &  Ti\,\textsc{ii} &     8710 &  31207 &  $-$0.690 \\   
$4468.500 \pm 1.30$  &  4468.500  &  Ti\,\textsc{ii} &     9118 &  31491 &  $-$0.270 \\   
$4501.273 \pm 1.30$  &  4501.273  &  Ti\,\textsc{ii} &     8998 &  31207 &  $-$0.684 \\   
$4522.634 \pm 1.30$  &  4522.634  &  Fe\,\textsc{ii} &    22939 &  45044 &  $-$2.119 \\   
$4534.400 \pm 2.50$  &  4533.239  &  Ti\,\textsc{i} &      6843 &  28896 &   0.563 \\  
                     &  4533.969  &  Ti\,\textsc{ii} &     9976 &  32025 &  $-$0.612 \\   
                     &  4534.778  &  Ti\,\textsc{i} &      6743 &  28788 &   0.376 \\  
                     &  4535.570  &  Ti\,\textsc{i} &      6661 &  28703 &   0.172 \\  
$4549.550 \pm 1.45$  &  4549.474  &  Fe\,\textsc{ii} &    22810 &  44785 &  $-$1.957 \\   
                     &  4549.617  &  Ti\,\textsc{ii} &    12775 &  34748 &  $-$0.110 \\   
$4555.893 \pm 1.30$  &  4555.893  &  Fe\,\textsc{ii} &    22810 &  44754 &  $-$2.281 \\   
$4563.761 \pm 1.30$  &  4563.761  &  Ti\,\textsc{ii} &     9851 &  31757 &  $-$0.795 \\   
$4571.968 \pm 1.30$  &  4571.968  &  Ti\,\textsc{ii} &    12677 &  34543 &  $-$0.209 \\   
$4629.339 \pm 1.30$  &  4629.339  &  Fe\,\textsc{ii} &    22637 &  44233 &  $-$2.379 \\   
$4670.407 \pm 1.30$  &  4670.407  &  Sc\,\textsc{ii} &    10945 &  32350 &  $-$0.518 \\   
$4920.000 \pm 5.00$  &  4915.233  &  Ti\,\textsc{i} &     15220 &  35560 &  $-$0.945 \\  
                   &  4918.954  &  Fe\,\textsc{i} &     33507 &  53831 &  $-$0.672 \\ 
                   &  4918.993  &  Fe\,\textsc{i} &     23111 &  43435 &  $-$0.365 \\ 
                   &  4919.867  &  Ti\,\textsc{i} &     17424 &  37744 &  $-$0.260 \\ 
                   &  4920.502  &  Fe\,\textsc{i} &     22846 &  43163 &   0.058 \\ 
                   &  4921.769  &  Ti\,\textsc{i} &     17540 &  37852 &  $-$0.005 \\ 
                   &  4922.827  &  Sc\,\textsc{i} &     16023 &  36331 &  $-$0.418 \\ 
                   &  4923.927  &  Fe\,\textsc{ii}&     23318 &  43621 &  $-$1.319 \\ 
$4981.732 \pm 1.30$  &  4981.732  &  Ti\,\textsc{i} &      6843 &  26911 &   0.586 \\  
$5701.545 \pm 1.00$  &  5701.545  &  Fe\,\textsc{i} &     20641 &  38175 &  $-$1.565 \\  
$5705.464 \pm 1.00$  &  5705.466  &  Fe\,\textsc{i} &     34692 &  52214 &  $-$1.581 \\  
$5732.000 \pm 1.30$  &  5731.762  &  Fe\,\textsc{i} &     34329 &  51771 &  $-$1.174 \\  
                     &  5732.275  &  Fe\,\textsc{i} &     40257 &  57698 &  $-$1.191 \\  
$5747.954 \pm 1.00$  &  5747.954  &  Fe\,\textsc{i} &     37163 &  54555 &  $-$0.599 \\  
$5861.108 \pm 1.00$  &  5861.107  &  Fe\,\textsc{i} &     34547 &  51604 &  $-$2.761 \\  
$5934.654 \pm 1.00$  &  5934.653  &  Fe\,\textsc{i} &     31686 &  48532 &  $-$1.192 \\  
$5952.800 \pm 1.00$  &  5952.716  &  Fe\,\textsc{i} &     32134 &  48928 &  $-$2.513 \\  
                   &  5952.889  &  Fe\,\textsc{i} &     34040 &  50833 &  $-$3.725 \\ 
$6151.617 \pm 1.00$  &  6151.617  &  Fe\,\textsc{i} &     17550 &  33802 &  $-$3.582 \\  
$6165.360 \pm 1.00$  &  6165.361  &  Fe\,\textsc{i} &     33413 &  49628 &  $-$1.667 \\  
$6170.500 \pm 1.00$  &  6170.504  &  Fe\,\textsc{i} &     38678 &  54880 &  $-$0.654 \\  
                   &  6171.006  &  Fe\,\textsc{i} &     38175 &  54376 &  $-$1.788 \\ 
$6173.334 \pm 1.00$  &  6173.341  &  Fe\,\textsc{i} &     17927 &  34122 &  $-$3.081 \\  
                   &  6173.642  &  Fe\,\textsc{i} &     35856 &  52050 &  $-$3.413 \\ 
$6191.500 \pm 1.00$  &  6191.558  &  Fe\,\textsc{i} &     19621 &  35768 &  $-$1.287 \\  
$6210.658 \pm 1.30$  &  6210.658  &  Sc\,\textsc{i} &         0 &  16097 &  $-$1.090 \\  
$6265.132 \pm 1.00$  &  6265.141  &  Fe\,\textsc{i} &     17550 &  33507 &  $-$2.834 \\  
$6305.657 \pm 1.30$  &  6305.657  &  Sc\,\textsc{i} &       168 &  16023 &  $-$0.950 \\  
$6318.000 \pm 1.00$  &  6318.018  &  Fe\,\textsc{i} &     19788 &  35612 &  $-$2.338 \\  
$6336.823 \pm 1.00$  &  6336.830  &  Fe\,\textsc{i} &     29733 &  45509 &  $-$1.260 \\  
$6408.000 \pm 1.00$  &  6407.643  &  Fe\,\textsc{i} &     32874 &  48476 &  $-$3.620 \\  
                   &  6408.026  &  Fe\,\textsc{i} &     29733 &  45334 &  $-$1.230 \\ 
                   &  6408.332  &  Fe\,\textsc{i} &     35379 &  50980 &  $-$3.563 \\ 
$6475.624 \pm 1.00$  &  6475.632  &  Fe\,\textsc{i} &     20641 &  36079 &  $-$3.070 \\  
$6481.870 \pm 1.00$  &  6481.878  &  Fe\,\textsc{i} &     18378 &  33802 &  $-$3.080 \\  
$6807.000 \pm 1.00$  &  6806.622  &  Fe\,\textsc{i} &     44023 &  58710 &  $-$1.744 \\  
                   &  6806.843  &  Fe\,\textsc{i} &     21999 &  36686 &  $-$3.210 \\ 
                   &  6807.288  &  Fe\,\textsc{i} &     42533 &  57219 &  $-$2.735 \\ 
$6810.262 \pm 1.00$  &  6810.262  &  Fe\,\textsc{i} &     37158 &  51837 &  $-$1.120 \\  
$8434.957 \pm 1.30$  &  8434.957  &  Ti\,\textsc{i} &      6843 &  18695 &  $-$0.886 \\  
$8514.400 \pm 1.90$  &  8514.072  &  Fe\,\textsc{i} &     17727 &  29469 &  $-$2.229 \\
                   &  8515.109  &  Fe\,\textsc{i} &     24339 &  36079 &  $-$2.073 \\
$8518.300 \pm 1.50$  &  8518.028  &  Ti\,\textsc{i} &     17215 &  28952 &  $-$1.250 \\
                   &  8518.352  &  Ti\,\textsc{i} &     15157 &  26893 &  $-$1.089 \\
$8582.350 \pm 1.35$  &  8582.258  &  Fe\,\textsc{i} &     24119 &  35768 &  $-$2.133 \\
$8611.800 \pm 0.90$  &  8611.803  &  Fe\,\textsc{i} &     22947 &  34556 &  $-$1.900 \\
$8679.000 \pm 2.10$  &  8678.997  &  Fe\,\textsc{i} &     48516 &  60035 &  $-$3.806 \\
                   &  8679.632  &  Fe\,\textsc{i} &     40052 &  51570 &  $-$1.512 \\
$8682.900 \pm 1.30$  &  8682.979  &  Ti\,\textsc{i} &      8492 &  20006 &  $-$1.941 \\
$8688.950 \pm 1.65$  &  8688.624  &  Fe\,\textsc{i} &     17550 &  29056 &  $-$1.212 \\
$8692.000 \pm 1.00$  &  8692.331  &  Ti\,\textsc{i} &      8437 &  19938 &  $-$2.295 \\
$8711.500 \pm 3.00$  &  8710.174  &  Mg\,\textsc{i} &     47841 &  59319 &  $-$1.550 \\
                   &  8710.392  &  Fe\,\textsc{i} &     39626 &  51103 &  $-$0.555 \\
                   &  8712.676  &  Mg\,\textsc{i} &     47844 &  59319 &  $-$1.670 \\
                   &  8713.188  &  Fe\,\textsc{i} &     23784 &  35257 &  $-$3.148 \\
$8730.750 \pm 0.95$  &  8730.497  &  Ti\,\textsc{i} &     27026 &  38477 &  $-$2.024 \\
$8735.250 \pm 1.75$  &  8734.712  &  Ti\,\textsc{i} &      8492 &  19938 &  $-$2.384 \\
                   &  8736.020  &  Mg\,\textsc{i} &     47957 &  59401 &  $-$0.690 \\
$8742.250 \pm 0.75$  &  8742.446  &  Si\,\textsc{i} &     47352 &  58787 &  $-$0.630 \\
$8757.200 \pm 1.60$  &  8757.187  &  Fe\,\textsc{i} &     22947 &  34363 &  $-$2.026 \\
$8792.850 \pm 1.35$  &  8793.342  &  Fe\,\textsc{i} &     37163 &  48532 &  $-$0.196 \\
$8806.000 \pm 2.70$  &  8806.756  &  Mg\,\textsc{i} &     35051 &  46403 &  $-$0.137 \\
                   &  8808.170  &  Fe\,\textsc{i} &     40405 &  51754 &  $-$1.109 \\
$8824.360 \pm 1.15$  &  8824.220  &  Fe\,\textsc{i} &     17727 &  29056 &  $-$1.540 \\
$8838.750 \pm 1.25$  &  8838.428  &  Fe\,\textsc{i} &     23052 &  34363 &  $-$1.980 \\
\end{longtable}

\begin{longtable}{c c r r r}
\caption{\label{tab:abundlines}Diagnostic lines used for our determination of element abundances.}\\
\hline
\hline
\noalign{\smallskip}
$\lambda_\mathrm{line}\,/\,{\AA}$  &  Ion  &  $E_\mathrm{low}\,/\,{\mathrm{cm}^{-1}}$ & $E_\mathrm{up}\,/\,{\mathrm{cm}^{-1}}$ & $\log gf$\\
\noalign{\smallskip}
\hline
\noalign{\smallskip}
\endfirsthead
\caption{continued.}\\
\hline\hline
\noalign{\smallskip}
$\lambda_\mathrm{line}\,/\,{\AA}$  &  Ion  &  $E_\mathrm{low}\,/\,{\mathrm{cm}^{-1}}$ & $E_\mathrm{up}\,/\,{\mathrm{cm}^{-1}}$ & $\log gf$\\
\noalign{\smallskip}
\hline
\endhead
\noalign{\smallskip}
\hline
\endfoot
5889.951 &   Na\,\textsc{i}  &      0 &  16973  &   0.101  \\
5895.924 &   Na\,\textsc{i}  &      0 &  16956  &  $-$0.197  \\
6696.185 &   Al\,\textsc{i}  &  32435 &  47365  &  $-$1.576  \\
6698.670 &   Al\,\textsc{i}  &  25438 &  40272  &  $-$1.960  \\
6698.673 &   Al\,\textsc{i}  &  25348 &  40272  &  $-$1.647  \\
6905.646 &   Al\,\textsc{i}  &  32435 &  46912  &  $-$1.287  \\
7083.969 &   Al\,\textsc{i}  &  32435 &  46548  &  $-$1.111  \\
7361.568 &   Al\,\textsc{i}  &  32435 &  46016  &  $-$0.903  \\
7835.309 &   Al\,\textsc{i}  &  32435 &  45195  &  $-$0.649  \\
7836.134 &   Al\,\textsc{i}  &  32437 &  45195  &  $-$0.494  \\
7836.134 &   Al\,\textsc{i}  &  32437 &  45195  &  $-$1.795  \\
8773.896 &   Al\,\textsc{i}  &  32437 &  43831  &  $-$0.161  \\
8773.898 &   Al\,\textsc{i}  &  32437 &  43831  &  $-$1.462  \\
7725.046 &   S\,\textsc{i}  &   9239 &  22180  &  $-$6.000  \\
7698.974 &   K\,\textsc{i}  &      0 &  12985  &  $-$0.170  \\
5857.451 &   Ca\,\textsc{i}  &  23652 &  40720  &   0.257  \\
6161.297 &   Ca\,\textsc{i}  &  20349 &  36575  &  $-$1.293  \\
6162.173 &   Ca\,\textsc{i}  &  15316 &  31539  &  $-$0.167  \\
6169.042 &   Ca\,\textsc{i}  &  20349 &  36555  &  $-$0.804  \\
6169.563 &   Ca\,\textsc{i}  &  20371 &  36575  &  $-$0.527  \\
6343.308 &   Ca\,\textsc{i}  &  35819 &  51579  &   0.845  \\
6361.786 &   Ca\,\textsc{i}  &  35897 &  51611  &   0.954  \\
6449.810 &   Ca\,\textsc{i}  &  20335 &  35835  &  $-$0.550  \\
6455.600 &   Ca\,\textsc{i}  &  20349 &  35835  &  $-$1.350  \\
6462.570 &   Ca\,\textsc{i}  &  20349 &  35819  &   0.310  \\
6471.668 &   Ca\,\textsc{i}  &  20371 &  35819  &  $-$0.680  \\
6493.788 &   Ca\,\textsc{i}  &  20335 &  35730  &   0.140  \\
6499.654 &   Ca\,\textsc{i}  &  20349 &  35730  &  $-$0.650  \\
8498.023 &   Ca\,\textsc{ii} &  13650 &  25414  &  $-$1.312  \\
8542.091 &   Ca\,\textsc{ii}  &  13711 &  25414  &  $-$0.362  \\
8662.141 &   Ca\,\textsc{ii}  &  13650 &  25192  &  $-$0.623  \\
4729.200 &   Sc\,\textsc{i}  &  11520 &  32659  &  $-$0.502  \\
4729.236 &   Sc\,\textsc{i}  &  11558 &  32697  &  $-$0.385  \\
4734.105 &   Sc\,\textsc{i}  &  11520 &  32637  &  $-$0.110  \\
4753.161 &   Sc\,\textsc{i}  &      0 &  21033  &  $-$1.658  \\ 
4779.348 &   Sc\,\textsc{i}  &    168 &  21086  &  $-$1.613  \\ 
4791.511 &   Sc\,\textsc{i}  &    168 &  21033  &  $-$2.075  \\
5700.164 &   Sc\,\textsc{i}  &  11558 &  29096  &   0.290  \\
5717.307 &   Sc\,\textsc{i}  &  11610 &  29096  &  $-$0.505  \\
5724.107 &   Sc\,\textsc{i}  &  11558 &  29023  &  $-$0.627  \\
6305.657 &   Sc\,\textsc{i}  &    168 &  16023  &  $-$0.950  \\
6378.807 &   Sc\,\textsc{i}  &      0 &  15673  &  $-$2.632  \\
6413.324 &   Sc\,\textsc{i}  &    168 &  15757  &  $-$2.677  \\
4314.080 &   Sc\,\textsc{ii}  &   4988 &  28161  &  $-$0.220  \\
4431.370 &   Sc\,\textsc{ii}  &   4883 &  27444  &  $-$1.830  \\
6279.740 &   Sc\,\textsc{ii}  &  12102 &  28021  &  $-$1.265  \\
6309.920 &   Sc\,\textsc{ii}  &  12074 &  27918  &  $-$1.630  \\
4455.320 &   Ti\,\textsc{i}  &  11640 &  34079  &   0.480  \\
4518.023 &   Ti\,\textsc{i}  &   6661 &  28788  &  $-$0.252  \\
4522.796 &   Ti\,\textsc{i}  &   6599 &  28703  &  $-$0.265  \\
4533.239 &   Ti\,\textsc{i}  &   6843 &  28896  &   0.563  \\
4534.778 &   Ti\,\textsc{i}  &   6743 &  28788  &   0.376  \\
4535.570 &   Ti\,\textsc{i}  &   6661 &  28703  &   0.172  \\
4535.916 &   Ti\,\textsc{i}  &   6599 &  28639  &  $-$0.026  \\
4536.043 &   Ti\,\textsc{i}  &   6557 &  28596  &  $-$0.129  \\
4548.765 &   Ti\,\textsc{i}  &   6661 &  28639  &  $-$0.274  \\
4552.456 &   Ti\,\textsc{i}  &   6743 &  28703  &  $-$0.262  \\
4656.468 &   Ti\,\textsc{i}  &      0 &  21469  &  $-$1.344  \\ 
4681.908 &   Ti\,\textsc{i}  &    387 &  21740  &  $-$1.129  \\
4981.732 &   Ti\,\textsc{i}  &   6843 &  26911  &   0.586  \\
6258.713 &   Ti\,\textsc{i}  &  11777 &  27750  &  $-$0.090  \\
8382.530 &   Ti\,\textsc{i}  &   6599 &  18525  &  $-$1.632  \\
8426.506 &   Ti\,\textsc{i}  &   6661 &  18525  &  $-$1.253  \\
4025.140 &   Ti\,\textsc{ii}  &   4898 &  29734  &  $-$1.960  \\
4394.060 &   Ti\,\textsc{ii}  &   9851 &  32603  &  $-$1.669  \\
4395.040 &   Ti\,\textsc{ii}  &   8744 &  31491  &  $-$0.660  \\
4417.720 &   Ti\,\textsc{ii}  &   9396 &  32026  &  $-$1.240  \\
4468.500 &   Ti\,\textsc{ii}  &   9118 &  31491  &  $-$0.270  \\
4549.617 &   Ti\,\textsc{ii}  &  12775 &  34748  &  $-$0.110  \\
4563.761 &   Ti\,\textsc{ii}  &   9851 &  31757  &  $-$0.795  \\
4352.870 &   V\,\textsc{i}  &    553 &  23520  &  $-$0.800  \\
4379.240 &   V\,\textsc{i}  &   2425 &  25254  &   0.600  \\
4384.706 &   V\,\textsc{i}  &    553 &  23353  &  $-$1.905  \\
4384.720 &   V\,\textsc{i}  &   2311 &  25112  &   0.000  \\
4395.230 &   V\,\textsc{i}  &   2153 &  24899  &   0.320  \\
4406.072 &   V\,\textsc{i}  &   8579 &  31268  &  $-$1.000  \\
4406.640 &   V\,\textsc{i}  &   2425 &  25112  &  $-$0.280  \\
4407.637 &   V\,\textsc{i}  &   2311 &  24993  &  $-$0.840  \\
4408.200 &   V\,\textsc{i}  &   2220 &  24899  &  $-$0.100  \\
4408.508 &   V\,\textsc{i}  &   2112 &  24789  &  $-$0.610  \\
4408.512 &   V\,\textsc{i}  &   2153 &  24830  &  $-$0.130  \\
4408.520 &   V\,\textsc{i}  &   2112 &  24789  &  $-$0.820  \\
4419.940 &   V\,\textsc{i}  &   2220 &  24839  &  $-$1.480  \\
4420.120 &   V\,\textsc{i}  &   2153 &  24771  &  $-$2.252  \\
4459.760 &   V\,\textsc{i}  &   2311 &  24728  &  $-$0.570  \\
4460.290 &   V\,\textsc{i}  &   2425 &  24839  &  $-$0.240  \\
5698.520 &   V\,\textsc{i}  &   8579 &  26122  &  $-$0.036  \\
5727.048 &   V\,\textsc{i}  &   8716 &  26172  &   0.088  \\
5737.059 &   V\,\textsc{i}  &   8579 &  26004  &  $-$0.675  \\
6135.361 &   V\,\textsc{i}  &   8476 &  24771  &  $-$0.750  \\
6150.157 &   V\,\textsc{i}  &   2425 &  18680  &  $-$1.277  \\
6243.105 &   V\,\textsc{i}  &   2425 &  18438  &  $-$0.878  \\
6274.649 &   V\,\textsc{i}  &   2153 &  18086  &  $-$1.657  \\
6531.440 &   V\,\textsc{i}  &   9825 &  25131  &  $-$1.320  \\
6531.466 &   V\,\textsc{i}  &  23935 &  39241  &  $-$2.931  \\
4274.800 &   Cr\,\textsc{i}  &      0 &  23386  &  $-$0.231  \\
4274.891 &   Cr\,\textsc{i}  &  24200 &  47586  &  $-$2.233  \\
4289.720 &   Cr\,\textsc{i}  &      0 &  23305  &  $-$0.360  \\ 
4527.332 &   Cr\,\textsc{i}  &  20524 &  42606  &  $-$0.906  \\
4535.695 &   Cr\,\textsc{i}  &  20524 &  42565  &  $-$0.570  \\
4600.741 &   Cr\,\textsc{i}  &   8095 &  29825  &  $-$1.305  \\
4600.775 &   Cr\,\textsc{i}  &  23934 &  45663  &  $-$2.354  \\
4652.152 &   Cr\,\textsc{i}  &   8095 &  29585  &  $-$1.026  \\
4708.018 &   Cr\,\textsc{i}  &  25549 &  46783  &   0.110  \\ 
4718.426 &   Cr\,\textsc{i}  &  25771 &  46959  &   0.097  \\ 
4829.314 &   Cr\,\textsc{i}  &  20524 &  41225  &  $-$1.630  \\
4829.372 &   Cr\,\textsc{i}  &  20524 &  41225  &  $-$0.787  \\
5783.886 &   Cr\,\textsc{i}  &  26796 &  44081  &  $-$0.177  \\
5787.965 &   Cr\,\textsc{i}  &  26796 &  44069  &   0.033  \\
5791.006 &   Cr\,\textsc{i}  &  26788 &  44051  &   0.324  \\
6924.179 &   Cr\,\textsc{i}  &  27820 &  42258  &  $-$0.135  \\
7462.378 &   Cr\,\textsc{i}  &  23499 &  36896  &  $-$0.040  \\
8947.180 &   Cr\,\textsc{i}  &  25039 &  36212  &  $-$0.724  \\
4554.988 &   Cr\,\textsc{ii}  &  32837 &  54784  &  $-$1.249  \\
4235.295 &   Mn\,\textsc{i}  &  23297 &  46901  &  $-$0.030  \\
4458.260 &   Mn\,\textsc{i}  &  24788 &  47212  &  $-$0.042  \\
4761.512 &   Mn\,\textsc{i}  &  23819 &  44815  &  $-$0.138  \\ 
4762.367 &   Mn\,\textsc{i}  &  23297 &  44289  &   0.426  \\ 
4765.846 &   Mn\,\textsc{i}  &  23720 &  44696  &  $-$0.077  \\ 
4766.418 &   Mn\,\textsc{i}  &  23549 &  44523  &   0.098  \\ 
4823.524 &   Mn\,\textsc{i}  &  18705 &  39431  &   0.144  \\ 
6013.513 &   Mn\,\textsc{i}  &  24779 &  41404  &  $-$0.397  \\
4045.820 &   Fe\,\textsc{i}  &  11976 &  36686  &   0.280  \\
4063.605 &   Fe\,\textsc{i}  &  12561 &  37163  &   0.072  \\
4063.627 &   Fe\,\textsc{i}  &  33096 &  57698  &  $-$0.691  \\
4071.740 &   Fe\,\textsc{i}  &  12969 &  37521  &  $-$0.022  \\
4271.760 &   Fe\,\textsc{i}  &  11976 &  35379  &  $-$0.164  \\
4325.739 &   Fe\,\textsc{i}  &      0 &  23111  &  $-$4.815  \\
4325.760 &   Fe\,\textsc{i}  &  12969 &  36079  &  $-$0.300  \\
4383.550 &   Fe\,\textsc{i}  &  11976 &  34782  &   0.200  \\
4404.761 &   Fe\,\textsc{i}  &  12561 &  35257  &  $-$0.142  \\
5701.545 &   Fe\,\textsc{i}  &  20641 &  38175  &  $-$1.565  \\
4233.113 &   Fe\,\textsc{ii}  &  54871 &  78487  &  $-$3.448  \\
4233.137 &   Fe\,\textsc{ii}  &  74498 &  98115  &  $-$2.864  \\
4233.170 &   Fe\,\textsc{ii}  &  20831 &  44447  &  $-$1.995  \\
4555.893 &   Fe\,\textsc{ii}  &  22810 &  44754  &  $-$2.281  \\ 
4813.449 &   Co\,\textsc{i}  &  23153 &  43922  &  $-$2.121  \\
4813.467 &   Co\,\textsc{i}  &  25938 &  46707  &   0.050  \\
6450.247 &   Co\,\textsc{i}  &  13796 &  29295  &  $-$1.698  \\
6814.942 &   Co\,\textsc{i}  &  15774 &  30444  &  $-$1.700  \\
7052.868 &   Co\,\textsc{i}  &  15774 &  29949  &  $-$1.440  \\
4519.979 &   Ni\,\textsc{i}  &  13521 &  35639  &  $-$2.570  \\
4715.757 &   Ni\,\textsc{i}  &  28578 &  49778  &  $-$0.331  \\ 
4786.531 &   Ni\,\textsc{i}  &  27580 &  48467  &  $-$0.244  \\
4831.169 &   Ni\,\textsc{i}  &  29084 &  49778  &  $-$0.291  \\
4918.362 &   Ni\,\textsc{i}  &  30980 &  51306  &  $-$0.109  \\
4984.112 &   Ni\,\textsc{i}  &  30619 &  50678  &   0.226  \\
5892.868 &   Ni\,\textsc{i}  &  16017 &  32982  &  $-$2.141  \\
6314.653 &   Ni\,\textsc{i}  &  15610 &  31442  &  $-$2.402  \\
6482.810 &   Ni\,\textsc{i}  &  15610 &  31031  &  $-$2.630  \\
6914.559 &   Ni\,\textsc{i}  &  15734 &  30192  &  $-$2.270  \\
7409.346 &   Ni\,\textsc{i}  &  30619 &  44112  &  $-$0.237  \\
7414.500 &   Ni\,\textsc{i}  &  16017 &  29501  &  $-$2.570  \\
5700.237 &   Cu\,\textsc{i}  &  13245 &  30784  &  $-$2.312  \\
5782.127 &   Cu\,\textsc{i}  &  13245 &  30535  &  $-$1.720  \\
4810.528 &   Zn\,\textsc{i}  &  32890 &  53672  &  $-$0.137  \\
7800.259 &   Rb\,\textsc{i}  &      0 &  12817  &   0.137  \\
7947.597 &   Rb\,\textsc{i}  &      0 &  12579  &  $-$0.167  \\
4741.918 &   Sr\,\textsc{i}  &  14318 &  35400  &  $-$0.320  \\
4872.488 &   Sr\,\textsc{i}  &  14504 &  35022  &  $-$0.200  \\
6504.000 &   Sr\,\textsc{i}  &  18067 &  33442  &   0.260  \\
7070.070 &   Sr\,\textsc{i}  &  14899 &  29039  &  $-$0.180  \\
4235.934 &   Y\,\textsc{i}  &    530 &  24131  &  $-$0.490  \\
4839.855 &   Y\,\textsc{i}  &  11532 &  32188  &   0.480  \\
6191.718 &   Y\,\textsc{i}  &      0 &  16146  &  $-$0.970  \\
6222.578 &   Y\,\textsc{i}  &      0 &  16066  &  $-$1.700  \\
6435.004 &   Y\,\textsc{i}  &    530 &  16066  &  $-$0.820  \\
4235.730 &   Y\,\textsc{ii}  &   1045 &  24647  &  $-$1.425  \\
4982.129 &   Y\,\textsc{ii}  &   8328 &  28394  &  $-$1.290  \\
7881.881 &   Y\,\textsc{ii}  &  14833 &  27517  &  $-$0.570  \\
4236.550 &   Zr\,\textsc{i}  &      0 &  23604  &  $-$1.000  \\
4772.323 &   Zr\,\textsc{i}  &   5023 &  25972  &   0.040  \\
4784.913 &   Zr\,\textsc{i}  &   5540 &  26434  &  $-$0.490  \\
5879.782 &   Zr\,\textsc{i}  &   1241 &  18244  &  $-$1.670  \\
6127.475 &   Zr\,\textsc{i}  &   1241 &  17556  &  $-$1.060  \\
6134.585 &   Zr\,\textsc{i}  &      0 &  16296  &  $-$1.280  \\
6143.252 &   Zr\,\textsc{i}  &    570 &  16844  &  $-$1.100  \\
6990.869 &   Zr\,\textsc{i}  &   5023 &  19324  &  $-$1.220  \\
7102.954 &   Zr\,\textsc{i}  &   5249 &  19324  &  $-$0.840  \\
8070.115 &   Zr\,\textsc{i}  &   5889 &  18277  &  $-$0.790  \\
4443.000 &   Zr\,\textsc{ii}  &  11984 &  34485  &  $-$0.160  \\
4205.303 &   Nb\,\textsc{i}  &   392  &   24165  & $-$0.850   \\
4523.397 &   Nb\,\textsc{i}  &  1143  &   23244  & $-$0.800   \\ 
4546.818 &   Nb\,\textsc{i}  &  1587  &   23574  & $-$0.750   \\ 
4573.075 &   Nb\,\textsc{i}  &  2154  &   24015  & $-$0.560   \\ 
4663.818 &   Nb\,\textsc{i}  &  1587  &   23023  & $-$0.740   \\ 
5751.408 &   Mo\,\textsc{i}  &  11454 &  28837  &  $-$1.014  \\
5791.839 &   Mo\,\textsc{i}  &  11454 &  28715  &  $-$1.046  \\
5858.266 &   Mo\,\textsc{i}  &  11858 &  28924  &  $-$0.996  \\
6619.134 &   Mo\,\textsc{i}  &  10768 &  25872  &  $-$1.252  \\
4031.626 &   Tc\,\textsc{i}  &   2573 &  27370  &   0.39   \\
4095.668 &   Tc\,\textsc{i}  &   3251 &  27660  &  $-$0.01   \\
4238.191 &   Tc\,\textsc{i}  &      0 &  23588  &  $-$0.39   \\
4262.270 &   Tc\,\textsc{i}  &      0 &  23455  &  $-$0.18   \\
4297.058 &   Tc\,\textsc{i}  &      0 &  23265  &  $-$0.03   \\
4206.015 &   Ru\,\textsc{i}  &   8084 &  31853  &  $-$0.480  \\
4385.385 &   Ru\,\textsc{i}  &   7483 &  30280  &  $-$0.610  \\
4385.645 &   Ru\,\textsc{i}  &   9058 &  31853  &  $-$0.490  \\
4410.025 &   Ru\,\textsc{i}  &   9184 &  31853  &  $-$0.380  \\
4460.027 &   Ru\,\textsc{i}  &   8771 &  31186  &  $-$0.530  \\
4554.517 &   Ru\,\textsc{i}  &   6545 &  28495  &   0.130  \\
4554.029 &   Ba\,\textsc{ii}  &      0 &  21952  &   0.170  \\
4934.076 &   Ba\,\textsc{ii}  &      0 &  20262  &  $-$0.150  \\
5853.668 &   Ba\,\textsc{ii}  &   4874 &  21952  &  $-$1.000  \\
6141.713 &   Ba\,\textsc{ii}  &   5675 &  21952  &  $-$0.076  \\
6496.930 &   Ba\,\textsc{ii}  &   4874 &  20262  &   0.130  \\
4354.400 &   La\,\textsc{ii}  &   7340 &  30305  &  $-$0.210  \\
4354.412 &   La\,\textsc{ii}  &   7395 &  30353  &  $-$0.500  \\
4526.111 &   La\,\textsc{ii}  &   6227 &  28315  &  $-$0.770  \\
4574.860 &   La\,\textsc{ii}  &   1394 &  23247  &  $-$1.140  \\
4662.498 &   La\,\textsc{ii}  &      0 &  21442  &  $-$1.240  \\
4970.386 &   La\,\textsc{ii}  &   2592 &  22705  &  $-$1.190  \\
5797.565 &   La\,\textsc{ii}  &   1971 &  19214  &  $-$1.410  \\
5805.773 &   La\,\textsc{ii}  &   1016 &  18236  &  $-$1.610  \\
6390.477 &   La\,\textsc{ii}  &   2592 &  18236  &  $-$1.450  \\
4324.785 &   Ce\,\textsc{ii}  &   7713 &  30829  &  $-$0.514  \\
4324.790 &   Ce\,\textsc{ii}  &   7662 &  30785  &  $-$0.050  \\
4386.827 &   Ce\,\textsc{ii}  &   1874 &  24663  &  $-$0.582  \\
4408.851 &   Ce\,\textsc{ii}  &   7234 &  29909  &  $-$0.965  \\
4408.870 &   Ce\,\textsc{ii}  &   7179 &  29860  &   0.120  \\
4408.894 &   Ce\,\textsc{ii}  &  10314 &  32989  &  $-$0.857  \\
4418.780 &   Ce\,\textsc{ii}  &   6968 &  29592  &   0.310  \\
4427.916 &   Ce\,\textsc{ii}  &   4323 &  26900  &  $-$0.460  \\
4427.920 &   Ce\,\textsc{ii}  &   4275 &  26859  &  $-$0.610  \\
4427.916 &   Ce\,\textsc{ii}  &   4323 &  26900  &  $-$0.460  \\
4427.920 &   Ce\,\textsc{ii}  &   4275 &  26859  &  $-$0.610  \\
4428.438 &   Ce\,\textsc{ii}  &   4266 &  26841  &  $-$0.657  \\
4444.700 &   Ce\,\textsc{ii}  &   8532 &  31024  &   0.110  \\
4483.893 &   Ce\,\textsc{ii}  &   6968 &  29263  &   0.010  \\
4483.900 &   Ce\,\textsc{ii}  &   6937 &  29239  &  $-$0.050  \\
4572.278 &   Ce\,\textsc{ii}  &   5514 &  27378  &   0.001  \\
4429.254 &   Pr\,\textsc{ii}  &   2998 &  25569  &  $-$0.010  \\
4205.600 &   Nd\,\textsc{ii}  &   5086 &  28857  &   0.070  \\
4232.380 &   Nd\,\textsc{ii}  &    513 &  24134  &  $-$1.020  \\
4351.290 &   Nd\,\textsc{ii}  &   1470 &  24445  &  $-$1.210  \\
4358.161 &   Nd\,\textsc{ii}  &   2585 &  25524  &  $-$0.280  \\
4358.170 &   Nd\,\textsc{ii}  &   4512 &  27449  &  $-$0.060  \\
4358.161 &   Nd\,\textsc{ii}  &   2585 &  25524  &  $-$0.280  \\
4358.170 &   Nd\,\textsc{ii}  &   4512 &  27449  &  $-$0.060  \\
4385.660 &   Nd\,\textsc{ii}  &   1650 &  24445  &  $-$0.550  \\
4391.100 &   Nd\,\textsc{ii}  &   2585 &  25352  &  $-$0.240  \\
4414.440 &   Nd\,\textsc{ii}  &    513 &  23160  &  $-$0.840  \\
4446.390 &   Nd\,\textsc{ii}  &   1650 &  24134  &  $-$0.500  \\
4680.737 &   Nd\,\textsc{ii}  &    513 &  21872  &  $-$1.260  \\
4706.543 &   Nd\,\textsc{ii}  &      0 &  21241  &  $-$0.880  \\
4715.586 &   Nd\,\textsc{ii}  &   1650 &  22850  &  $-$1.070  \\
4820.339 &   Nd\,\textsc{ii}  &   1650 &  22390  &  $-$1.240  \\
4229.713 &   Sm\,\textsc{ii}  &    327 &  23962  &  $-$1.224  \\
4390.855 &   Sm\,\textsc{ii}  &   1489 &  24257  &  $-$0.804  \\
4420.524 &   Sm\,\textsc{ii}  &   2689 &  25304  &  $-$0.695  \\
4433.890 &   Sm\,\textsc{ii}  &   3499 &  26046  &  $-$0.572  \\
4676.902 &   Sm\,\textsc{ii}  &    327 &  21702  &  $-$1.407  \\
4522.581 &   Eu\,\textsc{ii}  &   1669 &  23774  &  $-$0.678  \\
6645.064 &   Eu\,\textsc{ii}  &  11128 &  26173  &   0.204  \\
7426.569 &   Eu\,\textsc{ii}  &  10313 &  23774  &  $-$0.149  \\
4053.640 &   Gd\,\textsc{i}  &    999 &  25661  &   0.297  \\
4191.075 &   Gd\,\textsc{ii}  &   3444 &  27298  &  $-$0.653  \\
4394.720 &   Gd\,\textsc{ii}  &   6533 &  29288  &  $-$0.060  \\
4394.720 &   Gd\,\textsc{ii}  &   6605 &  29353  &  $-$1.783  \\
4752.526 &   Tb\,\textsc{ii}  &      0 &  21036  &  $-$0.816  \\
4186.819 &   Dy\,\textsc{i}  &      0 &  23878  &   0.693  \\
4077.966 &   Dy\,\textsc{ii}  &    828 &  25343  &  $-$0.058  \\
4409.383 &   Dy\,\textsc{ii}  &      0 &  22673  &  $-$1.420  \\
4301.596 &   Er\,\textsc{ii}  &      0 &  23241  &  $-$1.487  \\
4419.608 &   Er\,\textsc{ii}  &  13572 &  36192  &   0.386  \\
7131.816 &   Hf\,\textsc{i}  &      0 &  14018  &  $-$1.690  \\
7237.112 &   Hf\,\textsc{i}  &   4568 &  18382  &  $-$0.840  \\
4294.605 &   W\,\textsc{i}  &   2951 &  26230  &  $-$0.735  \\
4659.853 &   W\,\textsc{i}  &      0 &  21454  &  $-$1.900  \\
4260.848 &   Os\,\textsc{i}  &      0 &  23463  &  $-$1.440  \\
4420.468 &   Os\,\textsc{i}  &      0 &  22616  &  $-$1.530  \\
4793.993 &   Os\,\textsc{i}  &   4159 &  25013  &  $-$1.990  \\
\end{longtable}

\twocolumn

\end{document}